\documentclass[usenatbib]{mn2e}
\usepackage{epsfig}
\usepackage{natbib}
\usepackage{latexsym}
\usepackage{amsmath}
\usepackage{setspace}

\newcommand{\etal}{et~al.} 
\newcommand{\ionhy}{H{\sc ii} }

\newcommand{\kms}{$\mbox{km~s}^{-1}$}

\newcommand{\specsfig}[1]        
{
   \begin{center}
     \begin{minipage}[t]{0.45\textwidth}
         \psfig{file=#1.eps,height=0.9\textwidth,angle=270}
     \end{minipage}
     \end{center}
 }

\newcommand{\specdfig}[2]        
{
   \begin{center}
     \begin{minipage}[t]{0.45\textwidth}
         \psfig{file=#1.eps,height=0.9\textwidth,angle=270}
     \end{minipage}
     \hfill
     \begin{minipage}[t]{0.45\textwidth}
         \psfig{file=#2.eps,height=0.9\textwidth,angle=270}
     \end{minipage}
   \end{center}
}

\begin{document}

\title[Maser-based evolutionary schemes]{Testing maser-based evolutionary schemes: A new search for 37.7-GHz methanol masers}
\author[Ellingsen \etal]{S.\ P. Ellingsen,$^{1}$\thanks{Email: Simon.Ellingsen@utas.edu.au} S.\ L. Breen,$^{2}$ M.\ A. Voronkov,$^{2}$ J.\ R. Dawson$^{1}$\\
  \\
  $^1$ School of Mathematics and Physics, University of Tasmania, Private Bag 37, Hobart, Tasmania 7001, Australia\\
  $^2$ Australia Telescope National Facility, CSIRO, PO Box 76, Epping, NSW 1710, Australia}

 \maketitle
  
\begin{abstract}
We have used the Australia Telescope National Facility Mopra 22-m antenna to search for  37.7-GHz ($7_{-2} \rightarrow 8_{-1}E$) methanol masers towards a sample of thirty six class~II methanol masers.  The target sources are the most luminous class~II methanol masers not previously searched for this transition, with isotropic peak 12.2-GHz maser luminosity greater than 250~Jy\,kpc$^2$ and isotropic peak 6.7-GHz maser luminosity greater than 800~Jy\,kpc$^2$.  Seven new 37.7-GHz methanol masers were detected as a result of the search.  The detection rate for 37.7-GHz methanol masers towards a complete sample of all such class~II methanol maser sites south of declination -20$^\circ$ is at least  30 percent. The relatively high detection rate for this rare methanol transition is in line with previous predictions that the 37.7-GHz transition is associated with a late stage of the class~II methanol maser phase of high-mass star formation.  We find that there is a modest correlation between the ratio of the 6.7- and 37.7-GHz maser peak intensity and the 6.7- and 12.2-GHz maser peak intensity (correlation coefficient 0.63 in a log-log plot).  We detected one new 38.3-GHz ($6_{2} \rightarrow 5_{3}\mbox{A}^-$) methanol maser towards G\,$335.789+0.174$.  This is only the fourth source for which maser emission has been detected in this transition and it is the only one for which emission is not also observed in the 38.5-GHz $6_{2} \rightarrow 5_{3}\mbox{A}^+$ transition.
\end{abstract}


\section{Introduction}

The details of the processes through which high-mass stars form remains one of the enduring problems of modern astrophysics.  These stars form in clusters, deep within dense, dusty molecular clouds which absorb the radiation at most wavelengths, and their rapid evolution significantly complicates observational investigations.  Interstellar masers emit at centimetre and millimetre wavelengths, which are able to pass unattenuated through the surrounding natal material.  The close association of methanol masers with the early stages of high-mass star formation makes them one of the most useful signposts of such regions \citep{Ellingsen06}.

There are four common types of interstellar masers associated with young high-mass star formation regions: ground-state main-line OH masers at 1.6 GHz, 22-GHz water masers, class~II methanol masers (characterised by the 6.7- and 12.2-GHz transitions) and class~I methanol masers (characterised by the 44-GHz transition).  Many star formation regions show more than one of these different types of masers, each of which requires physical conditions within a specific range \citep[e.g.][]{Cragg+05}.  The most commonly observed maser transitions are those for which strong population inversion occurs for a relatively wide range of densities and temperatures.  It has long been speculated that the presence and absence of different maser transitions could be used to infer details of the physical conditions within these regions, and detailed modelling has been undertaken towards a small number of regions which show emission in a large number of different maser transitions \citep[e.g.][]{Cesaroni+91,Cragg+01,Sutton+01}.  

The recent completion of large, unbiased surveys for 6.7-GHz methanol \citep{Caswell+10,Green+10,Caswell+11,Green+12a}, and 22-GHz water masers \citep{Walsh+11}, combined with the existing data for 1.6-GHz OH masers \citep{Caswell98} means that it is now possible to undertake statistical investigations of the masers associated with high-mass star formation regions.   \citet{Ellingsen+07} suggested that it might be possible to use the presence and absence of interstellar masers to infer an evolutionary timeline for high-mass star formation regions.  \citet{Breen+10a} quantified the ``straw man'' model of \citeauthor{Ellingsen+07} through comparison of class~II methanol maser, 1.6-GHz OH maser and radio continuum properties.  The evolutionary phase for both water masers and class~I methanol masers remains poorly determined, and there is evidence that class~I masers may arise more than once during the high-mass star formation process \citep{Voronkov+10a,Chen+11,Cyganowski+12}.  The results of the HOPS water maser survey should clarify the situation for water masers \citep{Walsh+11}, and an unbiased, medium to large area survey of class~I methanol masers is required to definitively determine their place within the scheme.

A number of studies \citep[e.g.][]{Breen+10a,Breen+11} have demonstrated that the luminosity of 6.7- and 12.2-GHz class~II methanol masers increases for more evolved regions (i.e. those associated with ground-state OH masers or centimetre-wavelength radio continuum emission).  They also find that the properties of the maser emission, (which are confined to a relatively small volume close to the high-mass young stellar object), are more sensitive tracers of the evolutionary state than mid-infrared colours, or other physical quantities which arise from larger scale processes.  The methanol molecule is observed to exhibit a large number of rare, weak class~II masers, with 18 transitions currently identified \citep{Ellingsen+12}.  The presence of a less-common maser transition indicates atypical physical conditions in the associated regions, implying either an unusual source, or perhaps more likely, a relatively brief evolutionary phase in the high-mass star formation process.  Hence investigations of the rarer, weaker methanol maser transitions can potentially be used to identify short-lived phases and refine the time resolution of the maser-based evolutionary timeline.  

Most previous searches for the rarer (and generally weaker) class~II maser transitions have targeted the sources with the highest intensity in the 6.7- and 12.2-GHz transitions \citep[e.g.][]{Ellingsen+03,Ellingsen+04,Cragg+04}.  The observed intensity of a maser source will clearly depend upon both its luminosity and distance, but the assumption of the maser-based evolutionary timeline is that the presence or absence of the rarer maser transitions should depend on intrinsic properties of the exciting source, such as luminosity, age and mass.  Recently \citet{Ellingsen+11a} undertook a search for the 37.7-, 38.3- and 38.5-GHz class~II methanol maser transitions towards a diverse range of sources.  The target sources included all known southern 107-GHz methanol masers and all northern class II methanol masers with a 6.7-GHz maser peak flux density greater than 75~Jy.  This search resulted in the detection of 8 new 37.7-GHz methanol masers, and investigation of the source properties showed that all the detections were associated with high luminosity 12.2-GHz methanol masers (none of these new detections were from the northern, intensity-selected sample) and because of this \citet{Ellingsen+11a} suggested that the 37.7-GHz methanol masers are associated with a brief phase, just prior to the end of the class II maser phase of star formation.

\citet{Ellingsen+11a} found that all the known 37.7-GHz methanol masers were associated with 12.2-GHz methanol maser sources with a peak (isotropic) luminosity in excess of 250 Jy\,kpc$^2$ and a peak luminosity for the 6.7-GHz transition in excess of 800 Jy\,kpc$^2$.  They also found that approximately 50 percent of sources meeting these criteria which had been searched for 37.7-GHz methanol masers showed a detection.  Since the 6.7- and 12.2-GHz methanol maser luminosity has been shown to be related to the age of the associated high-mass young stellar object, then the maser-based evolutionary timeline predicts 37.7-GHz methanol masers should be detected towards a large fraction of the most luminous class~II methanol masers, however, to date it is primarily the highest intensity of the high luminosity sources that have been searched.  We have used the methanol maser multibeam catalogue of 6.7-GHz methanol masers \citep{Caswell+10,Caswell+11,Green+10,Green+12a} and the accompanying 12.2-GHz maser catalogue \citep{Breen+12a,Breen+12b} to identify all class~II methanol masers south of declination $-20^{\circ}$ which meet the criteria of 12.2-GHz peak luminosity greater than 250~Jy\,kpc$^2$ and 6.7-GHz peak luminosity greater than 800~Jy\,kpc$^2$.  For the distance determination we used published parallax \citep{Reid+09} or HISA \citep{Green+McClure11} results where available, or the near kinematic distance calculated using the method of \citet{Reid+09} where it is not. We identified thirty six southern class~II methanol masers meeting these criteria which have not previously been searched for 37.7-GHz methanol masers.  Here we report observations towards these sources.  

The methanol multibeam survey is complete for 6.7-GHz methanol masers with a peak flux density of 1~Jy or greater \citep{Green+09a}, so is sensitive to all sources with a luminosity greater than 800~Jy\,kpc$^2$ to a distance of more than 28~kpc (i.e. the entire Galaxy).  The 12.2-GHz search of \citeauthor{Breen+12a} has a 5-$\sigma$ sensitivity better than 0.8 Jy for the majority of sources, which means that it is sensitive to all sources with a luminosity greater than 250~Jy\,kpc$^2$ to a distance of approximately 17.5~kpc (i.e. the vast majority of the Galaxy where we expect high-mass star formation to occur).  So the sample of the most luminous 6.7- and 12.2-GHz methanol masers we have identified is expected to be essentially complete for all sources in the Galaxy (subject to the caveat that improved distance determinations for some sources in the future many change their calculated luminosity).  These observations represent the first direct observational test of the predictions of the maser-based evolutionary timeline for high-mass star formation of \citet{Ellingsen+07} and \citet{Breen+10a}.

\section{Observations} \label{sec:observations}

The observations were carried out with the Australia Telescope National Facility (ATNF) Mopra 22m radio telescope during 2012 January 15-19.  The observations were made with the 7mm receiver system during a Director's time allocation.  The system temperature during the observations varied between 70 and 115 K, and was less than 80 K for the majority of sources (68 percent).  The Mopra spectrometer (MOPS) was configured with 14 IF bands (``zooms'') spread over the frequency range from 33.1 to 40.0 GHz, the same setup as used in the observations of \citet{Ellingsen+11a}.  Each IF band covered 138 MHz, with 4096 spectral channels per band and two orthogonal linear polarisations were recorded.  This configuration yields a velocity coverage of approximately 1000~\kms\/ with a velocity resolution of 0.32~\kms\/ (for a channel spacing of 0.27~\kms) for unsmoothed spectra.  The Mopra telescope has RMS pointing errors of less than 10 arcseconds and at 38 GHz the antenna has a half-power beam width of 73 arcseconds \citep{Urquhart+10}.

The thermal and maser-lines observed in the current search are the same as those covered in the earlier Mopra observations.  Here we report only the observations of the three class~II methanol maser transitions, the $7_{-2} \rightarrow 8_{-1}\mbox{E}$, $6_{2} \rightarrow 5_{3}\mbox{A}^-$ and  $6_{2} \rightarrow 5_{3}\mbox{A}^+$ at 37.7, 38.3 and 38.5 GHz respectively.  The observations of the $4_{-1} \rightarrow 3_{0}\mbox{E}$ class~I methanol maser at 36.2~GHz and the thermal lines in both the current and earlier search will be reported elsewhere.  We adopted rest frequencies of 37.703696, 38.293292 and 38.452652 GHz for the $7_{-2} \rightarrow 8_{-1}E$, $6_{2} \rightarrow 5_{3}A^-$ and $6_{2} \rightarrow 5_{3}A^+$ transitions respectively \citep{Xu+97}.

The observations of each source consisted of a series of position-switched integrations of approximately 60 seconds duration, with reference observations offset from the target position by 5 arcminutes in declination.  The data were processed using the ASAP (ATNF Spectral Analysis Package).  On-source integration times ranged from 230 to 1700 seconds, with most observations being of 290 seconds duration or longer. The system temperature was measured by a continuously switched noise diode.  At a frequency of 38~GHz the Mopra telescope has a main beam efficiency of approximately 0.52, which implies a scaling factor of 14 Jy K$^{-1}$ \citep{Urquhart+10}.  The RMS noise level in the final spectra (after averaging all observations for a particular source over both linear polarisations, but with no smoothing) varied between 0.16 and 0.60 Jy, and was less than 0.40 Jy for the majority of sources (65 percent).  The line-width of many of the 37.7-GHz methanol masers detected by \citet{Ellingsen+11a} is comparable to the spectral resolution of the observations, hence smoothing the spectra with a Hanning window (or other smoothing function) does not yield any benefit in detecting weak sources.  The estimated zenith opacity during the observations varied very little and was typically 0.07, which implies attenuation of around 10 percent.  The data have not been corrected for the affects of atmospheric absorption and taking into account the antenna pointing accuracy, flux density calibration and variations in opacity we estimate the measured flux densities to be accurate to 15 percent.

The primary sample for our study were the thirty six southern class II methanol maser sources with peak 12.2-GHz methanol maser luminosity $>$ 250 Jy\,kpc$^2$ and peak 6.7-GHz methanol maser luminosity $>$ 800 Jy\,kpc$^2$ not previously observed in the 37.7-GHz ($7_{-2} \rightarrow 8_{-1}E$) transition.  These sources are labeled as sample A in Table~\ref{tab:sources}.  We also re-observed all the southern 37.7-GHz masers detected by \citet{Ellingsen+11a}, (11 sources) and made more sensitive observations of a number of the non-detections (6 sources).  These sources are labeled as sample B in Table~\ref{tab:sources}.  In addition, we made sensitive observations of one non-methanol maser source G\,$240.316+0.071$ (labeled as sample C in Table~\ref{tab:sources}), which shows maser emission in the main-line ground-state OH, as well as the 4750- and 6035-MHz excited OH transitions \citep{Caswell98,Caswell97,Dodson+02}.

\section{Results}

\begin{figure*}
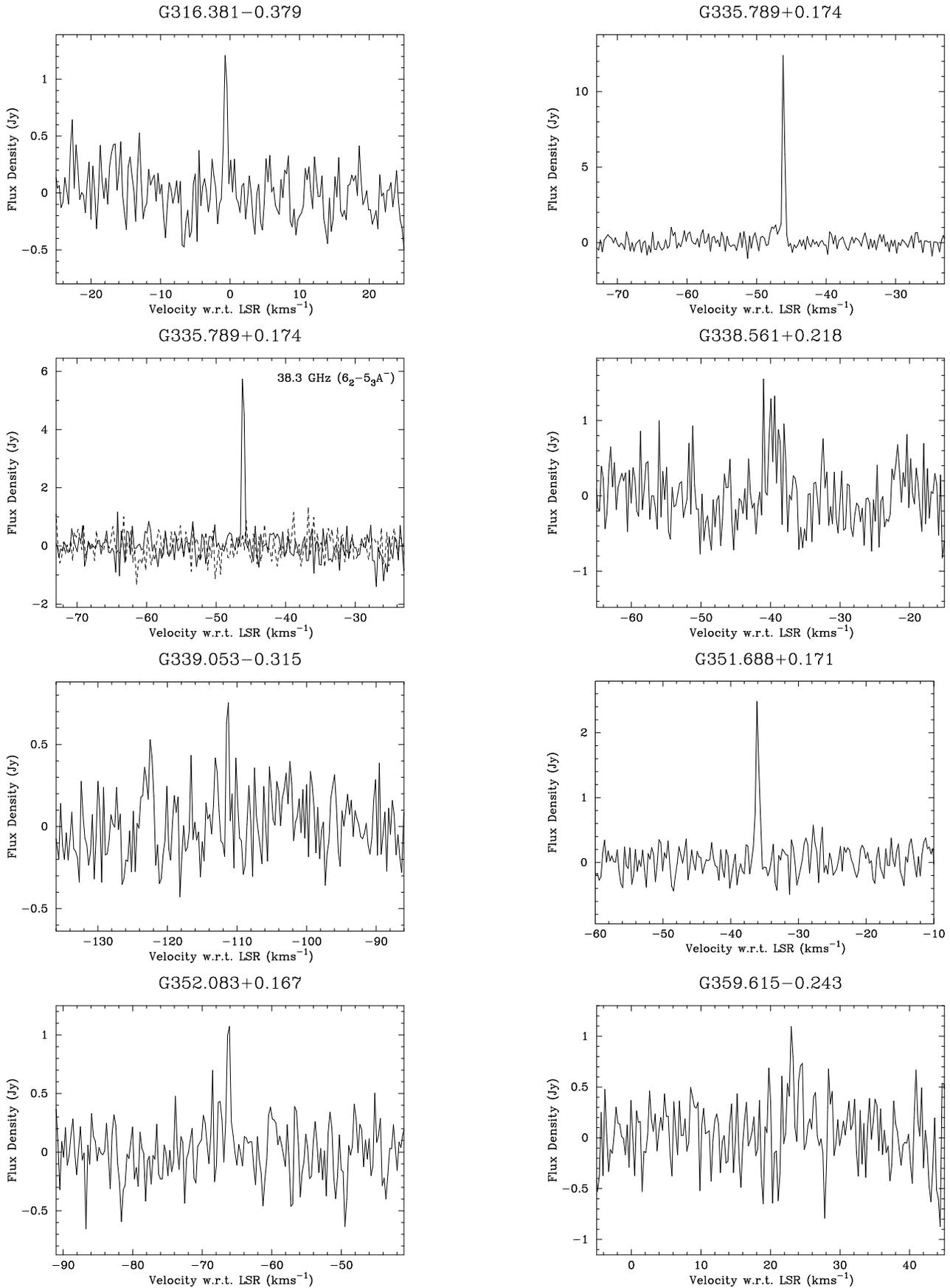

  \specdfig{G316.381-0.379_37ghz}{G335.789+0.174_37ghz}
  \specdfig{G335.789+0.174_383ghz}{G338.561-0.218_37ghz}
  \specdfig{G339.053-0.315_37ghz}{G351.688+0.171_37ghz}
  \specdfig{G352.083+0.167_37ghz}{G359.615-0.243_37ghz}
   \caption{Spectra of the newly detected 37.7- and 38.3-GHz methanol masers.  With the exception of the labelled spectrum of G\,$335.789+0.174$, all spectra are of the 37.7-GHz transition.  The dashed line in the 38.3-GHz spectrum of G\,$335.789+0.174$ shows the spectrum of the 38.5-GHz transition in this source.}
  \label{fig:newdet}
\end{figure*}

New 37.7-GHz methanol masers were detected towards 7 of the 36 sources in sample A, the spectra of the new detections are shown in Figure~\ref{fig:newdet}.  This corresponds to a detection rate of 19 percent in this sample.  The RMS noise level for each of the three class~II methanol maser transitions observed is listed for each source in Table~\ref{tab:sources} and is less than 0.4~Jy for the majority, corresponding to a typical 5-sigma detection limit of 2.0~Jy.  These observations are approximately a factor of 2 more sensitive than the previous search of \citet{Ellingsen+11a} which primarily targeted class~II methanol masers with high peak flux densities in the 6.7- and 12.2-GHz transitions.  The majority of the southern sources in the \citeauthor{Ellingsen+11a} search were luminous class~II methanol masers.  This is similar to the primary target sample for the current observations, however, sample A is comprised mainly of more distant sources, hence we expect the observed flux density of any detected 37.7-GHz methanol masers to be less and it is desirable to have greater sensitivity to reduce potential bias.

A velocity range of 200~\kms\/ centred on the 6.7-GHz peak velocity was searched for each source, for each of the three observed class~II methanol maser transitions (for four sources G\,$316.359-0.362$, G\,$352.083+0.167$, G\,$354.615+0.472$,  and G\,$358.371-0.468$ the velocity range searched was reduced to 150~\kms\/ due to poor baseline stability).  In all cases the emission from the newly detected 37.7-GHz methanol masers is restricted to a velocity range of less than a few \kms\/ and lies within the emission range of the associated 12.2- and 6.7-GHz masers, in many cases aligning very closely with the peak velocity of the 12.2-GHz transition (see Section~\ref{sec:align12}).  The close relationship between the emission in different class~II methanol transitions can be effectively used as {\em a priori} knowledge and makes it possible to reliably identify 37.7-GHz methanol masers which have peak flux densities less than the usually applied statistical limit of 5 times the RMS noise level in the spectrum.  Section~\ref{sec:indiv} outlines the evidence as to why each of the weaker sources shown in Figure~\ref{fig:newdet} is considered a bone fide detection, and Appendix~\ref{sec:additional} contains additional data which demonstrates we have a statistically significant detection for all seven new 37.7-GHz methanol masers.

\begin{table*}
\caption{The class~II methanol masers searched for emission in the 37.7-GHz ($7_{-2} \rightarrow 8_{-1}E$) transition.}
  \begin{tabular}{lllcccccccc} \hline
      \multicolumn{1}{c}{\bf Source} & \multicolumn{1}{c}{\bf RA}  & \multicolumn{1}{c}{\bf Dec} & \multicolumn{1}{c}{\bf Velocity} & \multicolumn{3}{c}{\bf RMS} & \multicolumn{1}{c}{\bf logarithm} & \multicolumn{1}{c}{\bf Sample}  \\
      \multicolumn{1}{c}{\bf name}     & \multicolumn{1}{c}{\bf (J2000)} & \multicolumn{1}{c}{\bf (J2000)} &  \multicolumn{1}{c}{\bf range} & \multicolumn{1}{c}{\bf 37.7~GHz} & \multicolumn{1}{c}{\bf 38.3~GHz} & \multicolumn{1}{c}{\bf 38.5~GHz} & \multicolumn{1}{c}{\bf 12.2-GHz peak}
       & \\
  & \multicolumn{1}{c}{\bf $h$~~~$m$~~~$s$}& \multicolumn{1}{c}{\bf $^\circ$~~~$\prime$~~~$\prime\prime$} & \multicolumn{1}{c}{\bf (km s$^{-1}$)} & \multicolumn{1}{c}{\bf (Jy)} & \multicolumn{1}{c}{\bf (Jy)} & \multicolumn{1}{c}{\bf (Jy)} & \multicolumn{1}{c}{\bf luminosity} & \\   \hline \hline   
 G\,$188.946+0.886$ & 06 08 53.3 & +21 38 29 & -90 -- 110 & 0.35 & 0.35 & 0.33 & 3.01 & B \\
 G\,$240.316+0.071$ & 07 44 51.9 & $-$24 07 42 & -40 --160 & 0.28 & 0.27 & 0.28 & - & C \\
 G\,$287.371+0.644$ & 10 48 04.4 & $-$58 27 01 & -100 -- 100 & 0.20 & 0.21 & 0.20 & 3.12 & A \\
 G\,$300.969+1.148$ & 12 34 53.4 & $-$61 39 40 & -140 -- 60 & 0.17 & 0.19 & 0.18 & 0.97 & B  \\
 G\,$309.921+0.479$ & 13 50 41.8 & $-$61 35 10 & -160 -- 40 & 0.47 & 0.46 & 0.47 & 3.61& B  \\
 G\,$310.144+0.760$ & 13 51 58.5 & $-$61 15 40 & -160 -- 40 & 0.16 & 0.15 & 0.16 & 3.17 & B \\
 G\,$311.643-0.380$ & 14 06 38.8 & $-$61 58 23 & -70 -- 130 & 0.27 & 0.28 & 0.26 & $<$0.89 & B \\
 G\,$313.469+0.190$ & 14 19 40.9 & $-$60 51 47 &  -110 -- 90 & 0.46 & 0.44 & 0.43 & 2.78 & A \\
 G\,$316.359-0.362$ & 14 43 11.2 &  $-$60 17 13 & -100 -- 100 & 0.22 & 0.23 & 0.23 & 3.18 & A \\
 G\,$316.381-0.379$  & 14 43 24.2 & $-$60 17 37 & -100 -- 100 & 0.22 & 0.23 & 0.22 & 2.42 & A \\
 G\,$317.466-0.402$  & 14 51 19.7 & $-$59 50 51 & -140 -- 60  & 0.44 & 0.44 & 0.45 & 2.76 & A \\
 G\,$318.948-0.196$ & 15 00 55.4 & $-$58 58 53 & -110 -- 40 & 0.28 & 0.28 & 0.28 & 4.20 & B \\
 G\,$323.740-0.263$ & 15 31 45.6 & $-$56 30 50 & -150 -- 50 & 0.45 & 0.46 & 0.44 & 3.50 & B\\
 G\,$326.475+0.703$ & 15 43 16.6 & $-$54 07 15 & -140 -- 60   & 0.47 & 0.47 & 0.48 & 3.38 & A \\
 G\,$327.402+0.445$ & 15 49 19.5 & $-$53 45 14 & -180 -- 20   & 0.45 & 0.46 & 0.47 & 3.00 & A \\
 G\,$329.029-0.205$ & 16 00 31.8 & $-$53 12 50 &  -140 -- 60   & 0.46 & 0.48 & 0.46 & 3.35 & A \\
 G\,$331.278-0.188$ & 16 11 26.6 &  $-$51 41 57 & -180 -- 20   & 0.48 & 0.45 & 0.46 & 3.17 & A \\
 G\,$331.425+0.264$ & 16 10 09.4 & $-$51 16 05 &  -190 -- 10 & 0.43 & 0.45 & 0.46 & 2.85 & A \\
 G\,$331.442-0.187$  & 16 12 12.5 & $-$51 35 10 & -190  -- 10 & 0.43 & 0.47 & 0.43 & 2.81 & A \\
 G\,$333.562-0.025$  & 16 21 08.8 & $-$49 59 48 & -140 -- 60   & 0.42 & 0.44 & 0.44 & 3.37 & A \\
 G\,$335.556-0.307$  & 16 30 56.0 & $-$48 45 50 &  -215 -- 15 & 0.23 & 0.24 & 0.23 & 2.69 & A \\
 G\,$335.726+0.191$ & 16 29 27.4 & $-$48 17 53 & -145 -- 55   & 0.43 & 0.43 & 0.45 & 3.08 & A \\ 
 G\,$335.789+0.174$ & 16 29 47.3 & $-$48 15 52 & -150 -- 50   & 0.42 & 0.43 & 0.43 & 3.23 & A \\
 G\,$336.358-0.137$  & 16 33 29.2 & $-$48 03 44 & -175 -- 25 & 0.41 & 0.44 & 0.42 & 2.50 & A \\
 G\,$337.201+0.114$ & 16 35 46.6 & $-$47 16 17 &  -160 -- 40 & 0.41 & 0.43 & 0.41 & 2.79 & A \\
 G\,$337.705-0.053$ & 16 38 29.7 & $-$47 28 03 & -155 -- 45 & 0.30 & 0.33 & 0.31 & 4.15 & B\\
 G\,$338.561+0.218$ & 16 40 38.0 & $-$46 11 26 & -140 -- 60   & 0.39 & 0.41 & 0.42 & 3.64 & A \\
 G\,$338.926+0.634$ & 16 40 13.6 & $-$45 38 33 & -165 -- 35  & 0.39 & 0.41 & 0.42 & 2.71 & A \\
 G\,$339.053-0.315$ & 16 44 49.0 & $-$46 10 13 &  -210 -- -10  & 0.17 & 0.18 & 0.17 & 3.67 & A \\
 G\,$339.762+0.054$ & 16 45 51.6 & $-$45 23 33 & -150 -- 50 & 0.39 & 0.41 & 0.38 & 2.59 & A \\
 G\,$339.884-1.259$ & 16 52 04.8 & $-$46 08 34 & -140 -- 60 & 0.47 & 0.46 & 0.47 & 3.74 & B \\
 G\,$339.949-0.539$  & 16 49 08.0 & $-$45 37 58 & -200 -- 0 & 0.40 & 0.40 & 0.40 & 2.52 & A \\
 G\,$339.986-0.425$ & 16 48 13.9 & $-$45 31 51 &  -190 -- 10   &  0.23 & 0.24 & 0.24 & 3.88 & A \\ 
 G\,$340.785-0.096$ & 16 50 14.8 & $-$44 42 25 & -210 -- 10 & 0.22 & 0.22 & 0.23 & 3.68 & B \\
 G\,$342.484+0.183$ & 16 55 02.3 & $-$43 13 00 & -140 -- 60   & 0.37 & 0.39 & 0.37 & 3.13 & A \\
 G\,$345.010+1.792$ & 16 56 47.7 & $-$40 14 26 & -120 -- 80 & 0.40 & 0.40 & 0.40 & 3.06 & B\\
 G\,$346.480+0.221$ & 17 08 00.1 & $-$40 02 16 & -120 -- 80   & 0.38 & 0.39 & 0.40 & 3.26 & A \\
 G\,$347.863+0.019$ & 17 13 06.2 & $-$39 02 40 & -135 -- 65 & 0.39 & 0.40 & 0.38 & 2.44 & A \\
 G\,$348.703-1.043$ & 17 20 04.1 & $-$38 58 30 & -100 -- 100 & 0.24 & 0.25 & 0.25 & 2.76 & B \\
 G\,$348.884+0.096$ & 17 15 50.1 & $-$38 10 12 & -175 -- 25 & 0.37 & 0.39 & 0.38 & 2.61 & A \\
 G\,$350.344+0.116$ & 17 20 00.0  & $-$36 58 00 & -165 -- 35  & 0.44 & 0.45 & 0.46 & 2.97 & A \\
 NGC6334F                & 17 20 53.4 & $-$35 47 00 & -110 -- 90 & 0.45 & 0.48 & 0.47 & 3.62 & B \\
 G\,$351.688+0.171$ & 17 23 34.5 & $-$35 49 46 & -140 -- 60   & 0.21 & 0.23 & 0.21 & 3.49 & A \\
 G\,$352.083+0.167$ & 17 24 41.2 & $-$35 30 19 & -165 -- 35 & 0.21 & 0.22 & 0.22 & 2.53 & A \\
 G\,$353.410-0.360$  & 17 30 26.2 & $-$34 41 46 & -120 -- 80 & 0.26 & 0.26 & 0.27 & 2.46 & B \\
 G\,$354.496+0.083$ & 17 31 31.8 & $-$33 32 44 & -75 -- 125 & 0.25 & 0.25 & 0.33 & 2.46 & A \\
 G\,$354.615+0.472$ & 17 30 17.1 & $-$33 13 55 & -100 -- 50 & 0.30 & 0.29 & 0.33 & 2.68 & A \\
 G\,$358.371-0.468$ & 17 43 32.0 & $-$30 34 11 & -75 -- 75  & 0.32 & 0.34 & 0.34 & 3.42 & A \\
 G\,$359.615-0.243$  & 17 45 39.1 & $-$29 23 30 & --80 -- 120 & 0.31 & 0.33 & 0.33 & 2.51& A \\
 G\,$0.092-0.663$       & 17 48 25.9 & $-$29 12 06 & -75 -- 125 & 0.31 & 0.31 & 0.32 & 2.63 & A \\
 G\,$8.832-0.028$      &  18 05 25.7 & $-$21 19 25 & -100 -- 100 & 0.25 & 0.25 & 0.24 & 2.99 & A \\
 G\,$9.621+0.196$    & 18 06 14.8 & $-$20 31 32 & -100 -- 100 & 0.60 & 0.60 & 0.59 & 4.04 & B\\
 G\,$23.440-0.182$    & 18 34 40.4 & $-$09 00 27  & 0 -- 200     & 0.30 & 0.31 & 0.27 & 2.49 & B \\
 G\,$35.201-1.736$   & 19 01 45.5 & +01 13 36 & -55 --145 & 0.26 & 0.25 & 0.23 & 3.07 & B \\ \hline
\end{tabular} \label{tab:sources}
\end{table*}

The data for the 38.3- and 38.5-GHz transitions, collected simultaneously with the 37.7-GHz were analysed in identical fashion.  The only new detection in either transition was towards G\,$335.789+0.174$, for which maser emission was detected in the 38.3-GHz $6_{2} \rightarrow 5_{3}A^-$ transition with a peak flux density of approximately 6~Jy, while there is no emission in the 38.5-GHz $6_{2} \rightarrow 5_{3}A^+$ transition stronger than a 3-$\sigma$ limit of 1.3~Jy (see Figure~\ref{fig:newdet}).  There are only three other sources which have been observed to show maser emission in either the 38.3- or 38.5-GHz transitions, W3(OH), NGC6334F and G\,$345.010+1.792$ \citep{Haschick+89,Ellingsen+11a}.  In all of these sources, both transitions are observed and typically have intensities within a factor of 2 of each other, and where multiple components are visible in the spectra they show very similar relative intensities in the two transitions.  On this basis the two transitions were considered a single transition group by \citet{Ellingsen+12}, however, in G\,$335.789+0.174$ the observed ratio of the two transitions exceeds a factor of 5.  The observation of the two transitions in G\,$335.789+0.174$ took place simultaneously, and observations of NGC6334F and G\,$345.010+1.792$ made during the same session detected emission in the 38.5-GHz $6_{2} \rightarrow 5_{3}A^+$ transition.  So it does not appear that any technical or observational error can account  for the observed difference in the relative strength of the 38.3- and 38.5-GHz emission in G\,$335.789+0.174$.

In addition to searching for new 37.7-, 38.3- and 38.5-GHz methanol masers towards previously unsearched luminous 12.2-GHz methanol masers we also made observations in each of these transitions towards all previously known sources visible from Mopra (that is all sources with the exception of W3(OH)).  All sources were detected, with comparable intensity and spectral shape to those observed in the 2009 May/June observations of \citet{Ellingsen+11a}.  Figure~\ref{fig:olddet} shows the 2012 January spectra of these previously known 37.7-GHz methanol masers, with the 2009 May/June spectra of \citet{Ellingsen+11a} plotted on the same scale with a dashed line.  We discuss the variability in the 37.7-, 38.3- and 38.5-GHz methanol masers further in section~\ref{sec:variability}.  

We also made more sensitive observations towards a small number of the non-detections from \citet{Ellingsen+11a} which were in sidereal time ranges where there were few sources in the primary target sample.  No emission was detected in any of the class~II methanol maser transitions towards any of these sources (G\,$240.316+0.071$, G\,$300.969+1.148$, G\,$309.921+0.479$, G\,$310.144+0.760$, G\,$311.643-0.380$ \& G\,$353.410-0.360$). The current observations are a factor of 2--8 times (most a factor of 5 or more) more sensitive than the observations of \citet{Ellingsen+11a}, and so demonstrate that any 37.7-GHz methanol masers towards these sources are very weak, and that the detection rate is not predominantly determined by the sensitivity of the observations (see also Sections~\ref{sec:complete} \& \ref{sec:align12}).

\begin{figure*}
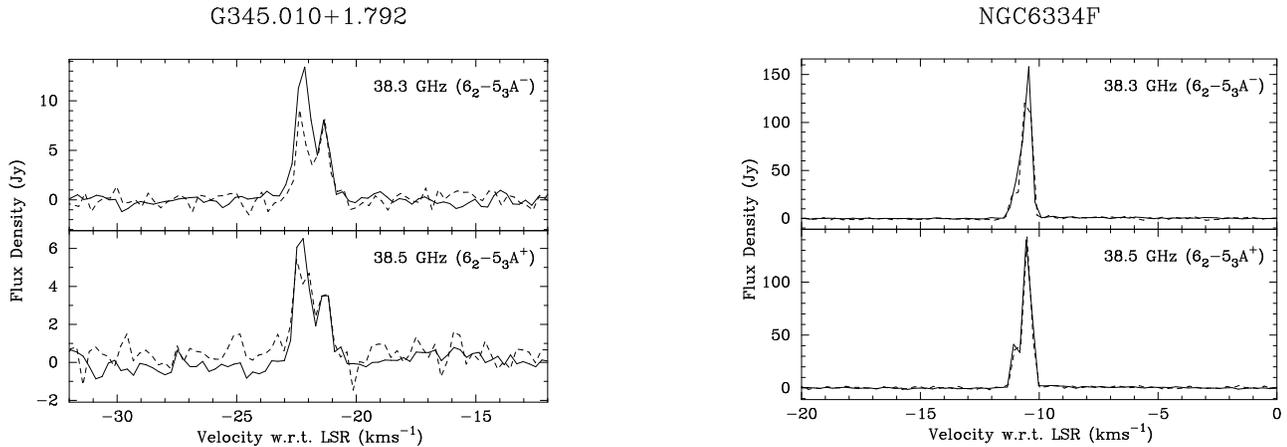

  \specdfig{G345.010+1.792_38}{NGC6334F_38}
   \caption{Spectra of previously detected 38.3- and 38.5-GHz methanol masers.  The current observations are shown as a solid line.  The dashed lines show the spectra from \citet{Ellingsen+11a}.  These were observed with the same correlator configuration and telescope in 2009 May/June.}
  \label{fig:old38}
\end{figure*}

Figure~\ref{fig:old38} shows the 2012 January spectra of the previously known 38.3- and 38.5-GHz methanol masers, with the 2009 May/June spectra of \citet{Ellingsen+11a} plotted on the same scale with a dashed line.  As for the 37.7-GHz methanol masers, the general shape and intensity of the spectra is similar between the two epochs.

We fitted one or more Gaussian profiles to the spectra of each of the detected maser sources in each transition.  Some sources require multiple Gaussian components to adequately (residuals comparable with the spectrum noise) describe the emission.  The results of these Gaussian fits are summarised in Tables~\ref{tab:gauss37} and \ref{tab:gauss38} for the 37.7- and the 38.3-/38.5-GHz transitions respectively.

\subsection{Comments on Individual Sources} \label{sec:indiv}

G\,$316.381-0.379$: The 37.7-GHz emission has a peak intensity 5.9 times the RMS noise level spectrum and lines up with the peak velocity of both the 6.7- and 12.2-GHz methanol masers in this source \citep{Green+12a,Breen+12b}.  The 12.2-GHz emission has a peak luminosity of 263 Jy kpc$^2$, only marginally above the cut-off for inclusion in sample~A.  The 12.2-GHz peak flux density reported for this source by \citet{Breen+12b} is 1.7~Jy, only a factor of 1.4 greater than the 37.7-GHz peak.  This is approximately an order of magnitude lower than typical 12.2-:37.7-GHz peak flux density ratios. \\

\noindent
G\,$335.789+0.174$: This is the strongest of the newly detected 37.7-GHz methanol masers, with a peak flux density of 12.4~Jy.  This source is also unusual in that it is the only new detection in the 38.3-GHz $6_{2} \rightarrow 5_{3}\mbox{A}^-$ transition.  The three previous sources (W3(OH), NGC6334F and G\,$345.010+1.792$) which have been observed as masers in this transition also show maser emission in the 38.5-GHz $6_{2} \rightarrow 5_{3}\mbox{A}^+$ line \citep{Ellingsen+11a}.  The peak flux densities in these two transitions are within a factor of 2 of each other in these sources, however, there is no hint of emission in the 38.5-GHz transition towards G\,$335.789+0.174$, with an RMS in the spectrum of 0.43 Jy.  We conservatively estimate the ratio of the 38.3-:38.5-GHz peak flux density to be $>$ 5. \\

\noindent
G\,$338.561-0.218$: The peak of the 37.7-GHz emission in this source is at 4.0 times the RMS noise level in the spectrum, but aligns in velocity with the 12.2-GHz emission seen in this source.  Additional observations undertaken in 2012 September with similar signal to noise confirm the detection (see Figure~\ref{fig:evidence}). \\

\noindent
G\,$339.053-0.315$: This is the weakest of the newly detected 37.7-GHz methanol masers, and the only one with a peak flux density $<$ 1 Jy.  The peak is 4.4 times the RMS noise level in the spectrum and aligns in velocity with the peak of the 12.2- and 6.7-GHz methanol masers in this source.  G\,$339.053-0.315$ has the second highest 12.2-GHz peak luminosity of the sources in sample~A, however, we were only able to detect the 37.7-GHz maser because the observation of this source was a factor of 3-4 times longer than for the majority of the sample.  Additional observations undertaken in 2012 September with similar signal to noise confirm the detection (see Figure~\ref{fig:evidence}). \\

\noindent
G\,$351.688+0.171$: This is the second strongest of the newly detected 37.7-GHz methanol masers and is associated with one of the most luminous 12.2-GHz methanol masers in sample~A. \\

\noindent
G\,$352.083+0.167$: This is another 37.7-GHz methanol maser which is associated with a 12.2-GHz methanol maser with a relatively low peak flux density \citep[2.8 Jy;][]{Breen+12a}. The peak emission is approximately 5 times the RMS noise level in the spectrum and aligns in velocity with the peak of the 6.7- and 12.2-GHz methanol masers in this source. \\

\noindent
G\,$359.615-0.243$: The 37.7-GHz methanol maser in this source is detected at 3.5 times the RMS noise level in the spectrum.  The 37.7-GHz peak lies within the velocity range of the 12.2-GHz emission, but is offset from the peak.  Additional observations undertaken in 2012 September with similar signal to noise confirm the detection (see Figure~\ref{fig:evidence}). \\

\noindent
G\,$35.201-1.736$ (W48): 37.7-GHz methanol maser emission with a peak flux density of approximately 20~Jy was detected in this source by \citet{Haschick+89}.  Onsala observations in 2005 detected a significantly weaker peak flux density of approximately 3~Jy and more recent observations with the Mopra telescope in 2009 and 2011 failed to detect 37.7-GHz emission in this source, with a 3-$\sigma$ sensitivity limit of 1.1 Jy \citep{Ellingsen+11a}.  The current observations (2012 January) do not detect any emission above a 3-$\sigma$ limit of 0.8~Jy.  This may be due to variability, but it is also possible (albeit unlikely), that over the 20 years since its initial detection that the 37.7-GHz methanol maser phase has finished in this source. We have included G\,$35.201-1.736$ as one of the 37.7-GHz detections in subsequent analysis in our discussion, as emission in this transition has been observed in this source, even though it is not present currently.

\begin{table*}
\caption{Characteristics of the sources detected in the 37.7 GHz methanol transition.  The formal uncertainty for each parameter from the fit is given in brackets.  Where that parameter was held fixed during fitting this is indicated with a --.  References to previous 37.7-GHz methanol maser observations are, 1=\citet{Haschick+89} ; 2 = \citet{Ellingsen+11a} ; sources marked with a $^*$ are newly detected 37.7-GHz methanol masers.}
  \begin{tabular}{lllrrrr} \hline
      \multicolumn{1}{c}{\bf Source} & \multicolumn{1}{c}{\bf RA}  & \multicolumn{1}{c}{\bf Dec} & \multicolumn{1}{c}{\bf Peak Flux} & \multicolumn{1}{c}{\bf Peak} & \multicolumn{1}{c}{\bf Full Width at} & {\bf References} \\
      \multicolumn{1}{c}{\bf name}     & \multicolumn{1}{c}{\bf (J2000)} & \multicolumn{1}{c}{\bf (J2000)} &  \multicolumn{1}{c}{\bf Density} & \multicolumn{1}{c}{\bf Velocity} & \multicolumn{1}{c}{\bf Half Maximum} & \\
  & \multicolumn{1}{c}{\bf $h$~~~$m$~~~$s$}& \multicolumn{1}{c}{\bf $^\circ$~~~$\prime$~~~$\prime\prime$} & \multicolumn{1}{c}{\bf (Jy)} & \multicolumn{1}{c}{\bf (\kms)} & \multicolumn{1}{c}{\bf (\kms)} \\   \hline \hline   G\,$188.946+0.886$ & 06 08 53.3 & +21 38 29 & 33.3(0.4) & 10.65(0.01) & 0.59(0.01) & 1,2 \\
G\,$316.381-0.379$ & 14 43 24.2 & $-$60 17 37 & 1.3(0.2)    & -0.63(0.04)   & 0.5(0.1) & * \\
G\,$318.948-0.196$ & 15 00 55.4 & $-$58 58 53 & 8.8(0.3)   & -34.21(0.01) & 0.47(0.02) & 2 \\
G\,$323.740-0.263$ & 15 31 45.6 & $-$56 30 50 & 21(1)     & -51.22(0.01) & 0.57(0.03) & 2 \\
				&		     &		         & 4.2(0.9) & -49.80(0.08) & 0.8(0.2)  & \\
G\,$335.789+0.174$ & 16 29 47.3 & $-$48 15 52 & 13.8(0.7) & -46.12(0.01) & 0.35(0.03) & * \\
                                      &                      &                   & 1.1(0.3) & -47.4(0.2)        & 1.3(0.4) &  \\
G\,$337.705-0.053$ & 16 38 29.7 & $-$47 28 03 & 3.3(0.3) & -54.93(0.02)   & 0.49(0.05) & 2 \\
G\,$338.561+0.218$ & 16 40 38.0 & $-$46 11 26 & 0.9(0.2) & -39.6(0.3)       & 2.2(0.6) & * \\
G\,$339.053-0.315$ & 16 44 49.0 & $-$46 10 13 & 0.7(0.2)  & -111.2(--)        & 0.6(0.2) & * \\
G\,$339.884-1.259$ & 16 52 04.8 & $-$46 08 34 & 406(5)  & -38.76(0.01)    & 0.31(0.01) & 2 \\
				&		     &		         & 74(5)    & -38.61(0.01)    & 0.64(0.01) & \\
				&		     &		         & 2.9(0.4) & -34.94(0.06)  & 2.4(0.2)   & \\
				&		     &		         & 2.4(0.4) & -33.66(0.07)  & 0.8(0.2)   & \\
G\,$340.785-0.096$ & 16 50 14.8 & $-$44 42 25 & 19.0(0.3) & -105.50(0.01) & 0.45(0.01) & 2\\
				&		      &	                  & 12.0(0.3) & -106.58(0.01) & 0.44(0.01) & \\
G\,$345.010+1.792$ & 16 56 47.7 & $-$40 14 26 & 203(2)  & -22.10(--) & 0.48(0.01) & 2 \\
	                             &                      &                   & 71(2)  & -21.83(--) & 0.72(0.02) & \\
	                             &                      &                    & 9(1)  & -20.9(--) & 0.7(0.1) & \\
G\,$348.703-1.043$ & 17 20 04.1 & $-$38 58 30 & 3.6(0.4)   & -3.51(0.02) & 0.40(0.05) & 2 \\
NGC6334F                & 17 20 53.4 & $-$35 47 00 & 105(2)  & -10.88(0.01)  & 0.32(0.01) & 1,2 \\
	                            &                       &                    & 27(1)   & -10.45(--)  & 0.36(0.01) & \\
	                             &                      &                    & 3.6(0.3) & -9.9(0.1)  & 3.3(0.2) & \\
G\,$351.688+0.171$ & 17 23 34.5 & $-$35 49 46 & 2.4(0.2)  & -36.06(0.02) & 0.57(0.06) & * \\
G\,$352.083+0.167$ & 17 24 41.2 & $-$35 30 19 & 1.2(0.2) & -66.21(0.05) & 0.5(0.1) & * \\
G\,$359.615-0.243$  & 17 45 39.1 & $-$29 23 30 & 0.6(0.1) & 23.4(0.3)       & 2.3(0.7) & * \\
G\,$9.621+0.196$    & 18 06 14.8 & $-$20 31 32 & 22(1) & -1.14(0.01) & 0.39(0.05) & 1,2 \\
                                      &                    &                    & 7.7(0.7) & -0.69(0.09) & 1.1(0.1) & \\
G\,$23.440-0.182$    & 18 34 40.4 & $-$09 00 27  & 1.1(0.2) & 98.16(0.08) & 0.9(0.2) & 2 \\ \hline
\end{tabular} \label{tab:gauss37}
\end{table*}

\begin{table*}
\caption{Characteristics of the sources detected in the 38.3- \& 38.5-GHz methanol transitions.  The formal uncertainty for each parameter from the fit is given in brackets.  Where that parameter was held fixed during fitting this is indicated with a --.  References to previous 38.3-GHz methanol maser observations are, 1=\citet{Haschick+89} ; 2 = \citet{Ellingsen+11a}, the source marked with a $^*$ is a newly detected 38.3-GHz methanol maser.}
  \begin{tabular}{llllrrrr} \hline
      \multicolumn{1}{c}{\bf Source} & \multicolumn{1}{c}{\bf RA}  & \multicolumn{1}{c}{\bf Dec} & \multicolumn{1}{c}{\bf Transition} &\multicolumn{1}{c}{\bf Peak Flux} & \multicolumn{1}{c}{\bf Peak} & \multicolumn{1}{c}{\bf Full Width at} & {\bf References} \\
      \multicolumn{1}{c}{\bf name}     & \multicolumn{1}{c}{\bf (J2000)} & \multicolumn{1}{c}{\bf (J2000)} & & \multicolumn{1}{c}{\bf Density} & \multicolumn{1}{c}{\bf Velocity} & \multicolumn{1}{c}{\bf Half Maximum} & \\
  & \multicolumn{1}{c}{\bf $h$~~~$m$~~~$s$}& \multicolumn{1}{c}{\bf $^\circ$~~~$\prime$~~~$\prime\prime$} & & \multicolumn{1}{c}{\bf (Jy)} & \multicolumn{1}{c}{\bf (\kms)} & \multicolumn{1}{c}{\bf (\kms)} \\   \hline \hline   
G\,$335.789+0.174$ & 16 29 47.3 & $-$48 15 52 & $6_{2} \rightarrow 5_{3}\mbox{A}^-$ & 7(1) & -46.12(0.01) & 0.36(0.06) & * \\
G\,$345.010+1.792$ & 16 56 47.7 & $-$40 14 26 & $6_{2} \rightarrow 5_{3}\mbox{A}^-$ & 13.8(0.4)  & -22.21(0.01) & 0.72(0.03) & 2 \\
	                             &                      &                       & $6_{2} \rightarrow 5_{3}\mbox{A}^-$ &  8.0(0.4)  & -21.31(0.02) & 0.48(0.04) & \\
G\,$345.010+1.792$ & 16 56 47.7 & $-$40 14 26 & $6_{2} \rightarrow 5_{3}\mbox{A}^+$ & 7.1(0.4)  & -22.29(0.02) & 0.64(0.04) & 2 \\
	                             &                      &                       & $6_{2} \rightarrow 5_{3}\mbox{A}^+$ & 4.0(0.4)  & -21.32(0.03) & 0.59(0.07) & \\
NGC6334F                & 17 20 53.4 & $-$35 47 00 & $6_{2} \rightarrow 5_{3}\mbox{A}^-$ & 170(10)  & -10.50(0.02)  & 0.35(0.02) & 1,2 \\
	                            &                       &                    & $6_{2} \rightarrow 5_{3}\mbox{A}^-$ & 40(2)   & -11.0(0.1)  & 0.4(0.1) & \\
NGC6334F                & 17 20 53.4 & $-$35 47 00 & $6_{2} \rightarrow 5_{3}\mbox{A}^+$ & 150(3)  & -10.49(0.01)  & 0.38(0.02) & 1,2 \\
	                            &                       &                    & $6_{2} \rightarrow 5_{3}\mbox{A}^+$ & 49(4)   & -11.01(0.02)  & 0.28(0.01) & \\ \hline
\end{tabular} \label{tab:gauss38}
\end{table*}

\begin{figure*}
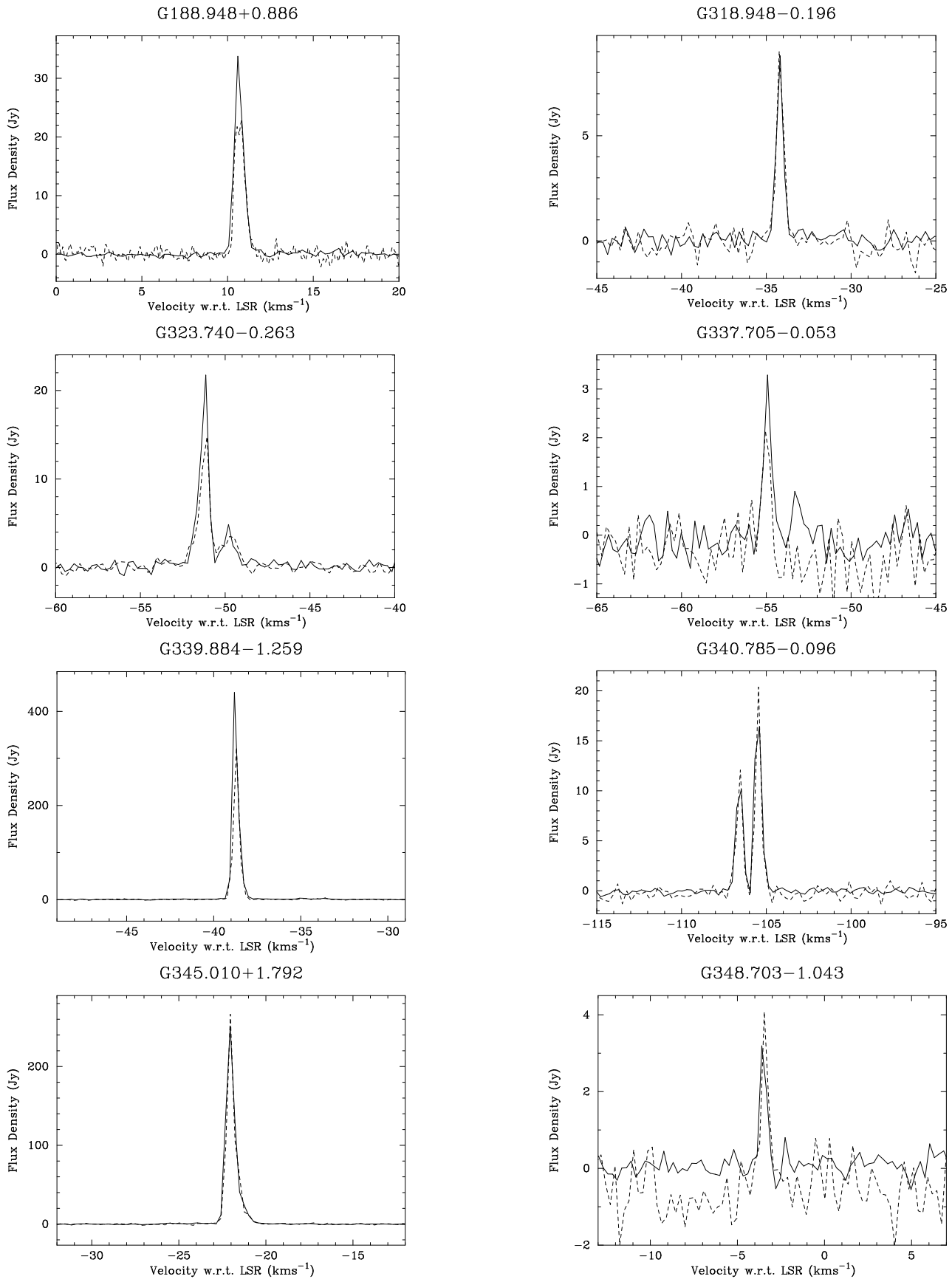

  \specdfig{G188.946+0.886_comp}{G318.948-0.196_comp}
  \specdfig{G323.740-0.263_comp}{G337.705-0.053_comp}
  \specdfig{G339.884-1.259_comp}{G340.785-0.096_comp}
  \specdfig{G345.010+1.792_comp}{G348.703-1.043_comp}
   \caption{Spectra of previously detected 37.7-GHz methanol masers.  The current observations are shown as a solid line.  The dashed line show the 37.7-GHz maser spectra from \citet{Ellingsen+11a}.  With the exception of G\,$188.948+0.886$, these were observed with the same correlator configuration and telescope in 2009 May/June.}
  \label{fig:olddet}
\end{figure*} 

\begin{figure*}
  \specdfig{NGC6334F_comp}{G9.621+0.196_comp}
  \specsfig{G23.440-0.182_comp}
   \contcaption{}
\end{figure*} 

\section{Discussion}

Prior to the current observations, a total of 79 class~II methanol maser sites had been searched for 37.7-GHz methanol masers \citep{Haschick+89,Ellingsen+11a}.  The 36 additional sources observed in the current work means that a total of 115 class~II methanol maser regions have been searched for this transition with detections towards 20 sources.  Previous searches for the 38.3- and 38.5-GHz transitions had covered 76 and 60 sources respectively \citep{Haschick+89,Ellingsen+11a}, the current search extends this to 112 and 96 sources respectively.  The total number of sources detected in these two transitions remains very low (4 and 3 respectively), indicating that they rarely occur towards luminous class~II methanol maser sites (and likely are rare in general).

To date, all reported observations of 37.7-GHz methanol masers have been made with single-dish telescopes with angular resolutions of order 1 arc minute.  In the discussion below we have assumed, that where emission in a rare or weak class~II methanol maser transition is detected towards a 6.7-GHz methanol maser, they arise from the same location (to within an arc second or less).  Although we cannot prove from these observations that there is coincidence between the 37.7-GHz (or 38.3-GHz) masers and the targeted 6.7-/12.2-GHz maser site, where high resolution observations have been made of different class~II methanol masers \citep[][]{Menten+92,Norris+93,Minier+00,Sutton+01}, this has always been found to be the case.  There are more than 300 known class~II methanol masers observed in transitions other than the 6.7-GHz, however, there is no evidence that any of these are not associated with a 6.7-GHz site.  The close relationship between the peak velocity of the 37.7- and 12.2-GHz methanol masers (see Section~\ref{sec:align12}), provides additional evidence that the assumption that all the detected 37.7-GHz methanol masers are associated with the targeted 6.7-GHz maser site is reasonable.  Future observations with the Australia Telescope Compact Array will enable the coincidence (or otherwise), of the 37.7-GHz transition with the other class~II masers to be determined conclusively.

\citet{Ellingsen+11a} found that all the sources towards which 37.7-GHz methanol masers had been detected had a 12.2-GHz methanol maser peak luminosity greater than 250 Jy\,kpc$^2$ and a 6.7-GHz methanol maser peak luminosity greater than 800~Jy\,kpc$^2$.  The primary aim of the current search for 37.7-GHz methanol masers was to test if the approximately 50\% detection rate for this transition towards the most luminous class~II methanol maser sources, reported by \citet{Ellingsen+11a}, is replicated across the entire population.  Of the thirty six sources south of declination -20$^{\circ}$ which had not previously been searched, we detected 37.7-GHz methanol masers towards 7, a detection rate of approximately 19 percent.  This is significantly less than that observed by \citet{Ellingsen+11a} and takes the detection rate for 37.7-GHz methanol masers over the entire sample of most luminous class~II masers to 31 percent.

The majority of targets for previous searches for the less common class~II methanol maser transitions have been the sources with the highest intensity in the 6.7- and 12.2-GHz transitions.  While these sources are typically also high luminosity within the overall population, selection based on intensity tends to exclude many sources with comparable luminosity at larger distances.  Here we are attempting to redress that bias, however, in primarily targeting more distant sources it is important to consider the effect that the sensitivity of the observations may have on our results.  The current observations have a sensitivity which is typically a factor of two or more better than the previous 37.7-GHz methanol maser search of \citet{Ellingsen+11a}, which has allowed us to identify a number of sources which would not have been detected at the sensitivity of the earlier observations.  

With a total sample of twenty 37.7-GHz class~II methanol masers we are able to compare the properties of the emission in this transition with that observed in the 6.7- and 12.2-GHz transitions.  Table~\ref{tab:fluxes} shows the peak flux density and velocity of the 37.7-GHz methanol masers compared to both the 6.7- and 12.2-GHz spectra.  Class~II methanol maser emission in most of these sources has been known for more than a decade and there are multiple observations of the 6.7- and 12.2-GHz masers reported in the literature.  We have selected the published 6.7- or 12.2-GHz peak flux densities and velocities from the observations which correspond most closely in time with the 37.7-GHz observations (this is generally the observations reported in the MMB catalogue papers and the 12.2-GHz catalogue papers of \citeauthor{Breen+12a}).  For this reason we have also used (where available) the 37.7-GHz peak flux densities and velocities from \citet{Ellingsen+11a}, as these data are closer in time to the comparison 6.7- and 12.2-GHz observations.  These data are used in the investigations of search completeness and the velocity alignment of the different methanol maser transitions that have been undertaken in sections~\ref{sec:complete} \& \ref{sec:align12}.

\begin{table*}
\caption{Peak flux density, and line-of-sight velocity of the peak flux density for the 6.7-, 12.2- and 37.7-GHz methanol masers towards all 37.7-GHz methanol masers.  Where possible we use the 6.7- and 12.2-GHz peak flux densities (and peak velocities) which are closest in time to the 37.7-GHz observation reported.  $R$ is the ratio of the 6.7-GHz peak flux density to the 1665 MHz OH maser peak flux density.  The 37.7-GHz references are * = this work ; 1 = \citet{Ellingsen+11a} ; 2 = \citet{Haschick+89}.  The 12.2-GHz references are 3 = \citet{Breen+12a} ; 4 = \citet{Breen+12b} ; 5 = \citet{Breen+10a} ; 6 = \citet{Caswell+95b} ; 7 = \citet{Menten+88a}.  The 6.7-GHz references are 8 = \citet{Caswell+10} ; 9 = \citet{Green+10} ; 10 = \citet{Caswell+11} ; 11 = \citet{Green+12a} ; 12 = \citet{Caswell+95a} ; 13 = \citet{Menten91b}.  The OH maser references are 14 = \citet{Reid+80} ; 15 = \citet{Cohen+88} ; 16 = James Green (pers. comm) ; 17 = \citet{Caswell98} ; 18 = \citet{Caswell+83} }
  \begin{tabular}{lrrlrrlrrlrl} \hline
      \multicolumn{1}{c}{\bf Source} & \multicolumn{3}{c}{\bf 37.7-GHz}  & \multicolumn{3}{c}{\bf 12.2-GHz} & \multicolumn{3}{c}{\bf 6.7-GHz} \\
      \multicolumn{1}{c}{\bf name}    & \multicolumn{1}{c}{\bf Flux} & \multicolumn{1}{c}{\bf Peak} & \multicolumn{1}{c}{\bf Ref.} & 
      \multicolumn{1}{c}{\bf Flux} & \multicolumn{1}{c}{\bf Peak} & \multicolumn{1}{c}{\bf Ref.} & \multicolumn{1}{c}{\bf Flux} &
      \multicolumn{1}{c}{\bf Peak} & \multicolumn{1}{c}{\bf Ref.} & \multicolumn{1}{c}{\bf R} & \multicolumn{1}{c}{\bf Ref.} \\
             & \multicolumn{1}{c}{\bf Density} & \multicolumn{1}{c}{\bf Velocity} &  & \multicolumn{1}{c}{\bf Density} & \multicolumn{1}{c}{\bf Velocity} & & 
       \multicolumn{1}{c}{\bf Density} & \multicolumn{1}{c}{\bf Velocity} &  \\ 
       & \multicolumn{1}{c}{\bf (Jy)} & \multicolumn{1}{c}{\bf (\kms)} &  & \multicolumn{1}{c}{\bf (Jy)} & \multicolumn{1}{c}{\bf (\kms)} & & 
       \multicolumn{1}{c}{\bf (Jy)} & \multicolumn{1}{c}{\bf (\kms)} &  \\ \hline
W3(OH)                      & 4.3       & -42.6    & 2 & 730   &   -44.8 & 7  &   3880  &  -44.2 & 13 & 14 & 14 \\ 
G\,$188.946+0.886$       & 23.4  & 10.7    & 1 &  225 &    10.9   & 4 &   607 &   10.8   & 11 & 1010 & 15\\   
G\,$316.381-0.379$  & 1.3     & -0.7     & *  &   1.7   &  -0.6  & 4  &     18   &   -0.5  & 11 & $>360$ & 16 \\     
G\,$318.948-0.196$   & 9.3     & -34.2  & 1 &  121 &   -34.5   & 4 &   570 &   -34.6  & 11 & 13 & 17 \\   
G\,$323.740-0.263$   & 14.6   & -51.2  & 1 &  396 &   -48.7   &4  &  3114 &   -50.5 & 11 & 1780 & 17\\  
G\,$335.789+0.174$  & 13.8  & -46.2   & *  &   162 &  -46.2 & 3 &   170  &  -47.5 & 10 & 260 & 17 \\     
G\,$337.705-0.053$   & 2.4     & -55.0  & 1 &  94    &  -54.6    & 3 &   171  &  -54.6  & 10 & 4.8 & 17 \\  
G\,$338.561+0.218$  & 0.9     & -39.6   & *  &   26    &  -40.2 & 3 &     38  &   -39.1 & 10 & $>760$ & 16 \\     
G\,$339.053-0.315$   & 0.7    & -111.2 & *  &   47    &  -111.8 & 3 &   130 &  -111.7 & 10 & 242 & 17 \\  
G\,$339.884-1.259$   & 323    & -38.7  & 1 &  846 &  -38.7    & 3 &   1520 &  -38.7 & 10 & 151 & 17 \\    
G\,$340.785-0.096$   & 20.5   & -105.5 & 1 &  42  & -105.3   & 3 &     158 &  -108.1 & 10 & 18 & 17 \\  
G\,$345.010+1.792$  &  207   & -22.1   & 1 &  296 &  -21.8   & 3 &    268   & -21.0    & 8 & 15 & 17 \\    
G\,$348.703-1.043$   & 4.4      & -3.4     & 1 & 34    &   -3.5    & 3 &     65     & -3.5      & 8 & 203 & 17 \\     
NGC6334F                   & 70.4   & -10.9   & 1 &  976 &   -10.4  & 3 &  3420   & -10.4   & 8 & 22 & 17 \\     
G\,$351.688+0.171$  & 2.4    & -36.1    & * &   21   &   -36.2  & 3 &    41  &   -36.1  & 8 & $>820$ & 16 \\        
G\,$352.083+0.167$  & 1.2    & -66.2    & * &  2.8   &   -66.2  &3 &    6.8   &   -66.0   & 8 & $>140$ & 16 \\       
G\,$359.615-0.243$   & 0.6    & 23.4     &  * & 5.7   &   19.3    & 3 &     39 &    19.3   & 8 & 8.6 & 17 \\       
G\,$9.621+0.196$       & 23.6   & -1.1     & 1 & 401  &   1.4      & 3 &  5240   & 1.3       & 9 & 526 & 17\\     
G\,$23.440-0.182$     & 2.3      & 98.0    & 1 &   9     &   97.5   & 5  &      77     & 103     & 12 & 16 & 18 \\  
G\,$35.201-1.736$     & 3.2      & 44.5    & 1 &  109  &   44.6  & 6  &  560   &  42      & 12  & 13 & 18 \\ \hline 
\end{tabular} \label{tab:fluxes}
\end{table*}

\subsection{The completeness of the 37.7-GHz search} \label{sec:complete}

\begin{figure}
   \begin{center}
     \begin{minipage}[t]{0.45\textwidth}
         \psfig{file=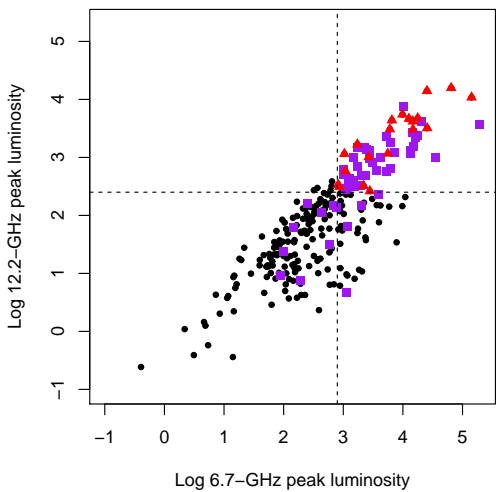,height=0.9\textwidth,angle=270}
     \end{minipage}
     \end{center}
   \caption{The 6.7-GHz peak luminosity versus the 12.2-GHz peak luminosity for the 12.2-GHz methanol masers detected by \citet{Breen+12a,Breen+12b} (black dots).  The sources with detected 37.7-GHz methanol masers are shown with red triangles, the sources which have been searched for 37.7-GHz methanol masers (either by \citet{Ellingsen+11a}, or the current observations), but have no detected emission are shown with purple squares.}
  \label{fig:lum}
\end{figure}

The effect of the sensitivity of the current observations on the completeness of the search is something that we would like to know.  It requires some means of estimating the likely intensity (or range of intensities) for 37.7-GHz methanol masers towards the target sources.  All of the detected 37.7-GHz methanol masers have associated 6.7- and 12.2-GHz methanol masers and we have collated their characteristics in Table~\ref{tab:fluxes}.  We have used the peak flux density for each of the three transitions (6.7, 12.2 and 37.7~GHz) to form three flux density ratios.  As flux density ratios are independent of distance estimates they are potentially useful in attempting to determine the completeness level of the current 37.7-GHz search. The ratio of the 12.2-:37.7-GHz peak flux densities and of the 6.7-:37.7-GHz peak flux densities vary over two and three orders of magnitude respectively, while the ratio of the 6.7-:12.2-GHz peak flux densities varies by approximately an order of magnitude for this sample.  For the 20 known 37.7-GHz methanol maser sources, the median flux density ratios are 11.3, 29.7 and 3.1 for the 12.2-:37.7-GHz, 6.7-:37.7-GHz and 6.7-:12.2-GHz respectively.  \citet{Caswell+95a} found that the median ratio for a large sample of 6.7-:12.2-GHz methanol masers is 5.4 \citep{Caswell+95b}.  \citet{Breen+11} showed that the luminosity of both 6.7- and 12.2-GHz methanol masers increases as the associated exciting source evolves, but found that the 6.7-GHz luminosity increases more rapidly.  So we might naively expect that if 37.7-GHz methanol masers are only associated with the most evolved class~II methanol masers then the median 6.7-:12.2-GHz flux density ratio for these should be larger than for the sample of all 12.2-GHz methanol masers (i.e. it should be greater than 5.4), but we find a median 6.7-:12.2-GHz ratio of 3.1 for our sample.  However, our results are consistent with those of \citet{Breen+11}, who found marginal evidence that for class~II methanol masers with an associated OH maser (which are thought to be older than those not associated with OH maser emission), the luminosity ratio decreases.  Figure~\ref{fig:lum} shows the 12.2-GHz peak luminosity versus the 6.7-GHz peak luminosity, with the 37.7-GHz methanol masers shown as red triangles and the sources searched for 37.7-GHz masers, but not detected shown as purple squares.  The 37.7-GHz methanol masers are clearly preferentially associated with those class~II methanol maser sites which have relatively more luminous 12.2-GHz methanol masers.

\begin{figure}
   \begin{center}
     \begin{minipage}[t]{0.45\textwidth}
         \psfig{file=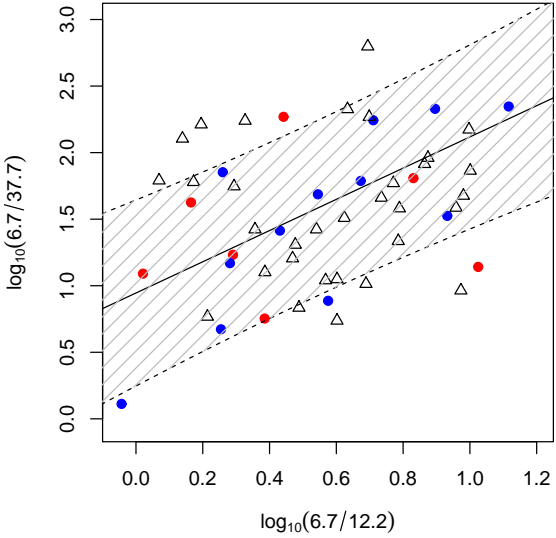,height=1.0\textwidth,angle=270}
     \end{minipage}
     \end{center}
   \caption{The ratio of the 6.7-:37.7-GHz peak flux density versus the 6.7-:12.2-GHz peak flux density for 37.7-GHz detections (filled circles), and non-detections (open triangles).  Blue circles are 37.7-GHz methanol masers known prior to the current search, red circles are the sources detected in the current work.  For the 37.7-GHz non-detections from the current observations we have plotted a lower limit to the 6.7-:37.7-GHz peak flux density ratio using 3 times the RMS noise level in the 37.7-GHz spectrum as an estimate of the upper limit of maser emission in that transition.  The line of best fit for the 37.7-GHz detections is plotted on the figure (solid line), the dashed lines and shading show the 68 percent confidence interval for the fit.}
  \label{fig:ratio}
\end{figure}

Plotting the 6.7-:37.7-GHz flux density ratio against the 6.7-:12.2-GHz flux density ratio for the 37.7-GHz methanol maser detections on a log-log scale (see Figure~\ref{fig:ratio}), we found the two quantities showed a modest correlation (correlation coefficient 0.63).  We have plotted the newly detected 37.7-GHz methanol masers using red circles and the previously known sources as blue circles, and it is clear that there are no significant differences between the two groups.  The solid line in this plot is a line of best fit to the 6.7-:37.7-GHz flux density ratio versus the 6.7-:12.2-GHz flux density ratio for the 37.7-GHz detections, so 50 percent of the known 37.7-GHz maser sources lie above this line.  We also investigated whether there is any relationship between the 12.2-:37.7-GHz and 6.7-:12.2-GHz flux density ratios, but found that they show no significant correlation (correlation coefficient 0.1) 

We can use the 6.7-:37.7-GHz versus the 6.7-:12.2-GHz flux density ratio relationship to assess the degree to which the detection rate of the current observations may be influenced by sensitivity. Using 3-$\sigma$ as an estimate of the upper limit of the 37.7-GHz flux density in the non-detected sources, we can find a lower limit to the 6.7-:37.7-GHz flux density ratio for these sources and we have plotted these as open triangles in Figure~\ref{fig:ratio}.  Nearly all the non-detected sources lie within the central 68 percent confidence interval of the fit (i.e. within 1-$\sigma$ of the mean for a normal distribution).  This suggests that the sensitivity of the current observations was sufficient to have a reasonable probability of detecting 37.7-GHz methanol masers in the majority of sources observed, but insufficient to be able to rule out the presence of weaker than average 37.7-GHz methanol masers in the non-detections.  There are 11 non-detections which lie above the solid line in Figure~\ref{fig:ratio}, meaning that the current lower limits on the 6.7-:37.7-GHz flux density ratio for these sources are higher than that observed for 50 percent or more of the detected 37.7-GHz methanol masers.  Assuming that the distribution observed for the 37.7-GHz methanol masers detected to date is representative of the population as a whole, then for these sources (particularly those with the highest lower limits), we can say that there is a low probability that they have an associated 37.7-GHz methanol maser.    We would also expect that some of the 21 non-detections below the solid line do have an associated 37.7-GHz methanol maser which is too weak to have been detected in the current observations.  However, it is not feasible to make any quantitative estimates of what fraction of sources that might be.  We further discuss the detection rate for 37.7-GHz methanol masers towards luminous class~II methanol maser sites, and the implications of the current observations as a test of maser-based evolutionary schemes in section~\ref{sec:evolution}.

\citet{Breen+12a} stacked the spectra of their 12.2-GHz non-detections at the velocity of the 6.7-GHz peak and from the absence of a peak in the stacked spectrum they concluded that there are few 12.2-GHz methanol masers with peak flux densities just below the detection limit.  We have undertaken a comparable analysis with the 37.7-GHz sample A non-detections from the current observations (a total of 30 sources), aligning them at the velocity of the 12.2-GHz peak (see section~\ref{sec:align12}).  The stacked spectrum has an RMS noise level of 70~mJy and shows no evidence for a population of weak 37.7-GHz methanol masers close to the detection threshold of the current observations.  

\subsection{Alignment between 37.7- and 12.2-GHz maser peaks} \label{sec:align12}

\citet{Breen+11} showed that in 80 percent of 12.2-GHz methanol masers the velocity of the 12.2- and 6.7-GHz peaks coincide.  Considering the peak velocities of the 12.2- and 37.7-GHz masers to be coincident if they differ in velocity by less than 0.6~\kms\/ (twice the velocity width of a spectral channel in the 37.7-GHz observations), we find that sixteen of twenty 37.7-GHz methanol maser peaks coincide with the 12.2-GHz peak (i.e. 80 percent).   Comparing the 37.7- and 6.7-GHz maser peak velocities we find eleven of twenty sources for which they coincide.  While the coincidence for the 12.2- and 37.7-GHz transitions is the same as that observed by \citeauthor{Breen+12a} for the 6.7- and 12.2-GHz transitions these figures are somewhat subjective as if we tighten the coincidence criteria to a velocity difference of 0.3~\kms\/ or less then our rates drop to 55 percent and 30 percent respectively.  

The methanol molecule has recently been shown to be unusually sensitive to variations in the proton-to-electron mass ratio $\mu$ \citep{Levshakov+11,Jansen+11}.  Variations in fundamental ``constants'' of physics, are of interest as they imply new physics, beyond the standard model of particle physics \citep[see][for a review of this area]{Uzan11}.  Different rotational transitions of methanol have different sensitivities to variations in $\mu$, and comparison of the frequency of different methanol rotational transitions in Galactic star formation regions with those measured in the laboratory can be used to place observational constraints on chameleon-like scalar fields.  

\citet{Ellingsen+11b} undertook a comparison of the peak velocity of 6.7- and 12.2-GHz methanol masers in a sample of 9 Galactic sources with simple spectra.  From these data they were able to place an upper limit on $|\frac{\Delta \mu}{\mu}| < 2.7 \times 10^{-8}$, while observations of the 9.9- and 104.0-GHz class~I methanol masers in {\em IRAS}16547$-$4247 by \citet{Voronkov+06} produce an upper limit of $|\frac{\Delta \mu}{\mu}| < 2.8 \times 10^{-8}$ \citep{Levshakov+11}.  Comparison of the frequencies of different astronomical maser transitions are potentially complicated by a range of factors \citep[see][for a detailed discussion]{Ellingsen+11b}, however all else being equal, a comparison between the 37.7-GHz transition and either the 12.2- or 6.7-GHz would be a factor of 3--4 more sensitive to variations in $\mu$ than previous investigations.  For some sources the differences between the 37.7- and 12.2-GHz methanol maser peak velocities observed here are almost certainly due to blending of different spectral components within the beam of the Mopra telescope, however, for those sources where the peak velocities are very similar, future high spatial and spectral resolution observations offer the prospect of more stringent tests of variations of $\mu$ at the current epoch.   The current observations have too coarse spectral resolution for it to be worthwhile undertaking such investigations with this data and in addition the positional accuracy 10 arcseconds is less than desirable for this type of comparison.  Observations of the 37.7-GHz transition with an interferometer, such as the ATCA offer the prospect of addressing both of those issues simultaneously and are the appropriate next step to further investigate this question.

\subsection{Variability of 37.7-GHz methanol masers} \label{sec:variability}


Figure~\ref{fig:olddet} shows a comparison of observations of previously known southern 37.7-GHz methanol masers made in 2012 January and 2009 May/June.  In most cases there are differences in the peak intensity of the observations at the two epochs, but the overall shape of the spectra are similar.  Gaussian fits for all the 2012 spectra in Figure~\ref{fig:olddet} are given in Table~\ref{tab:gauss37} and can be compared with the equivalent data for the 2009 spectra \citep[table~2 of][]{Ellingsen+11a}.  These show that the largest changes in the flux density are approximately 40 percent.  Both the 2009 and 2012 observations have an estimated accuracy of 15 percent for the flux density calibration, so the observed variations exceed the combined uncertainty by around 10 percent.  During the current observations some known 37.7-GHz methanol masers were observed more than once, and we can use those observations to test the estimated uncertainty in the flux density calibration.  Three observations of G\,$345.010+1.792$ showed variations in the peak flux density with a standard deviation of 11 percent, however, the standard deviation of the integrated intensity for the same observations was only 3.7 percent.  This suggests that the accuracy of the flux density calibration is within our 15 percent estimate, but that because the velocity resolution of the observations (0.32~\kms) is comparable with the width of many of the maser lines (typically 0.3--0.5~\kms) there is additional variation in the observed peak intensity due to under sampling in the spectral domain.  Comparing the current observations of 37.7-GHz methanol masers with those made by \citet{Ellingsen+11a} and \citet{Haschick+89}, it appears that in general 37.7-GHz methanol masers exhibit relatively little variation (less than 30 percent) on timescales of a few years, but can show more dramatic variability (e.g. G\,$35.201-1.736$ has varied by more than an order of magnitude), on timescales of tens of years.

\subsection{Associations with OH and water masers} \label{sec:OH}

The association of 37.7-GHz methanol masers with luminous 6.7- and 12.2-GHz methanol masers means that we would expect them to generally also exhibit OH maser emission \citep{Breen+10a}.  A search of the literature shows that 16 of the 20 known 37.7-GHz methanol masers have an associated 1.6-GHz OH maser \citep{Reid+80,Caswell+83,Cohen+88,Caswell98}.  There are no published OH maser data available for G\,$316.381-0.379$, G\,$338.561+0.218$, G\,$351.688+0.171$ or G\,$352.083+0.167$, however, these sources have been observed as part of the MAGMO project \citep{Green+12b}, and there is no ground state OH maser emission associated with any of these sources with a flux density limit of approximately 50 mJy (James Green personal communication).  \citet{Caswell97} found that sources with a high 6.7-GHz methanol to ground-state OH peak flux density ratio were less likely to have an associated ultra-compact H{\sc ii} region than if this ratio was low.  \citet{Caswell97} referred to this ratio as $R$, and grouped sources into three categories based on this.  Sources with $R > 32$ were called ``methanol-favoured'', while sources with $R<8$ were ``OH-favoured'' .  We have calculated $R$ for each of the 37.7-GHz methanol maser sources and the values are listed in the second last column of Table~\ref{tab:fluxes}.  For those sources where the OH data were taken from \citet{Caswell98}, we have scaled the OH maser flux density by a factor of 1.45 prior to calculating $R$ (as recommended by \citeauthor{Caswell98}, to account for the low spectral resolution of those observations).

Eleven of the twenty 37.7-GHz methanol maser sources are classified as methanol-favoured, with eight being intermediate (neither methanol- nor OH-favoured), and only one source (G\,$337.705-0.053$) is OH-favoured.  \citet{Caswell97} found that very few methanol-favoured sources have an associated strong (flux density greater than 100~mJy) ultra-compact \ionhy region, which is in contrast to the OH-favoured sources that often do.  He suggested that the methanol-favoured sources are younger than the OH-favoured sources.  Interestingly, the majority (6 of the 7) of the newly detected 37.7-GHz methanol masers are methanol-favoured sources.

The majority of the 37.7-GHz methanol maser detections have been searched for water maser emission, the exceptions being G\,$316.381-0.379$, G\,$351.688+0.171$ and G\,$352.083+0.167$.  The water maser search of \citet{Breen+10b} observed the 13 remaining sources south of declination -10$^\circ$, and detected water masers toward 11 of these sources (the non-detections were G\,$339.053-0.315$ and G\,$348.703-1.043$).  Water maser emission has been detected towards all the northern sources, G\,$188.946+0.886$ \citep{Batchelor+80}, G\,$23.440-0.182$ and G\,$35.201-1.736$ \citep{Forster+89} and W3(OH) \citep[][although the association here is arguable, as the water masers are offset by 6 arcseconds from the strong OH and methanol maser site]{Alcolea+93}.

The rate of association of both ground-state OH masers (80 percent) and water masers (88 percent) with the 37.7-GHz methanol maser detections is significantly higher than the rate for class~II methanol maser sources in general.  \citet{Caswell96} found ground-state OH masers associated with approximately 40 percent of class~II methanol masers from complete samples of both transitions, while \citet{Szymczak+05} found water masers associated with 52 percent of class~II methanol masers.  Summarising all of these results, we can characterise a typical 37.7-GHz methanol maser source as occurring in a region with luminous 6.7- and 12.2-GHz methanol masers, likely to have associated ground-state OH and water maser emission and in many cases radio continuum emission from an ultra-compact \ionhy region.  \citet{Ellingsen+11a} found that all the detected 37.7-GHz methanol masers have an associated 107-GHz methanol maser, so the newly detected sources are good candidates for future searches in this transition.

\subsection{Testing maser-based evolutionary schemes} \label{sec:evolution}

The primary aim of the current observations was to test the prediction of \citet{Ellingsen+11a} that 37.7-GHz methanol masers are associated with approximately 50 percent of the star formation regions with the most luminous 6.7- and 12.2-GHz methanol masers.  The 19 percent detection rate for the current search is well below that prediction.  Combining the results of \citet{Ellingsen+11a} with the current observations, all class~II methanol maser sites south of declination -20$^{\circ}$ with a 6.7-GHz methanol maser with an isotropic peak luminosity greater than 800~Jy\,kpc$^2$, and a 12.2-GHz peak luminosity greater than 250~Jy\,kpc$^2$ have now been searched for the 37.7-GHz transition.  In total there are 52 sources which meet these criteria and 16 of them are observed to have an associated 37.7-GHz methanol maser, corresponding to a detection rate towards this sample of 31 percent\footnote{\citet{Ellingsen+11a} stated that there were eight 37.7-GHz methanol masers observed towards 15 class~II maser sources meeting the luminosity criteria used for the current search.  This was incorrect, the correct figures for the \citet{Ellingsen+11a} search are 9 detections from 16 sources searched.  Combining these results with the 7 detections towards 36 sources observed in the current search gives a total of 16 detections from 52 sources searched.}.  We can use $\sqrt{N_{det}}$ as an estimate of the uncertainty in the number of detections $N_{det}$, in the population.  Hence, \citet{Ellingsen+11a} detected 37.7-GHz methanol masers towards 56 $\pm$ 19 percent of luminous class~II methanol maser sites, whereas combining their results with the current study we achieve a rate of 30.7 $\pm$ 7.7 percent.  These two estimates of the detection rate for 37.7-GHz methanol masers are consistent within the estimated uncertainties (just).  Although the difference between the detection rate initially achieved by \citet{Ellingsen+11a} and the rate we have determined for a complete sample may be due to chance, sensitivity may also play a role, as we show below.

We have investigated the completeness of our search for 37.7-GHz methanol masers in section~\ref{sec:complete} and demonstrated that it is likely that more sensitive observations will make some additional detections towards the current sample of sources.  We have also stacked the non-detection spectra at the velocity of the 12.2-GHz peak (see Section~\ref{sec:align12}), but see no sign of a peak in the stacked spectrum, which has an RMS of 70~mJy.  Given the relatively small number of non-detection spectra available to stack and the likely lower level of coincidence in velocity for the 37.7- and 12.2-GHz transitions than observed for the 12.2- and 6.7-GHz transitions, we cannot place any strong limits on the level of 37.7-GHz emission associated with the non-detected sources. However, if there were a number of 37.7-GHz methanol masers close to the detection threshold of the current observations we would expect to see evidence for this in the stacked spectrum.  The fact that we don't tells us that either there are very few sources, or that {\em on average} the peak flux density of these sources is well below the sensitivity of the current observations. From this we can be very confident that for any 37.7-GHz methanol masers associated with the non-detections the average flux density of these sources is less than 0.35~Jy (5 times the RMS noise level in the stacked spectrum).  This is approximately a factor of five more sensitive than the typical detection threshold for the individual observations.    Observations significantly more sensitive than those undertaken here could be undertaken with reasonable efficiency with an interferometer, as both the position and the velocity of the 37.7-GHz methanol masers can be predicted to a high degree of accuracy from the 6.7-GHz transition.  A thirty minute observation with the Australia Telescope Compact Array (ATCA) would yield sensitivities a factor of 5--10 higher than the current observations.

The previous 37.7-GHz methanol maser searches of \citet{Haschick+89} and \citet{Ellingsen+11a}, along with the current search have made observations towards 115 class~II methanol maser sites.  The 115 sources are comprised of the 52 sources from the complete sample of the most luminous class~II methanol masers south of declination -20$^\circ$ and 63 additional sources.  These 63 ``other'' sources are either a southern source (declination less than -20$^\circ$) which don't meet one, or both of the class~II maser luminosity criteria (20 sources), or a more northerly maser region.  The published observations of the more northerly 6.7- and 12.2-GHz methanol regions are both very heterogeneous in terms of the observing approach and the sensitivity, and also incomplete for the 12.2-GHz transition.  For that reason we have not attempted to estimate the 6.7- and 12.2-GHz peak luminosities for this sample.  Amongst the 63 ``other'' sources there are four detected 37.7-GHz methanol masers.  These are all in the sample of 43 sources north of declination -20$^\circ$, and each of the four detected 37.7-GHz methanol maser regions meet the luminosity criteria identified by \citet{Ellingsen+11a}.  If the 31 percent detection rate observed in the complete southern sample of luminous class~II methanol masers were characteristic of the population as a whole then we would expect approximately twenty 37.7-GHz methanol masers to have been detected towards these 63 ``other'' sources.

These results demonstrate that high luminosity 6.7- and 12.2-GHz methanol maser emission is a necessary condition for the presence of a 37.7-GHz methanol maser.  However, they also show that in excess of 50 percent of high-luminosity 6.7- and 12.2-GHz methanol masers do not have an associated 37.7-GHz methanol maser, indicating that it is not a sufficient condition.  An obvious question to ask is ``Is this consistent with the predictions of the maser-based evolutionary timeline of \citet{Ellingsen+07} and \citet{Breen+10a}?''  The premise of the maser-based evolutionary timeline for high-mass star formation is that statistically, the presence and absence of different maser transitions can be used to infer the relative evolutionary phase for the associated young stellar object.  In the absence of special geometry, masers will be seen by a particular observer to arise along lines of sight, where by chance, there is a sufficient degree of velocity coherence.  In this scenario we expect the luminosity of the masers will be correlated with the volume of gas where the conditions are suitable to produce a population inversion in the transition in question.  We would also predict that observers in different locations in the Galaxy will observe similar luminosities for a particular maser source, although different specific maser distributions.  The stochastic nature of the observed maser emission (chance velocity coherence for the specific line of sight), will clearly produce scatter in the expected correlation between gas volume and maser luminosity and the smaller the volume the greater this effect.  Water masers and class~I methanol masers are observed to be associated with outflows \citep{Genzel+81,Voronkov+06,Cyganowski+09}, which will clearly have a preferred direction and hence potentially constitute special geometry.  The outflows in general appear to have wide opening angles and the large number of both types of maser which are detected suggests that it is possible to detect maser emission for a large range of outflow orientations.  However, the influence of outflow geometry on observed maser luminosity is likely to cause additional scatter in relationships with exciting source properties for these types of masers.

Another potential complication for maser-based evolutionary schemes is the effect of metallicity.  Metallicity can affect molecular abundances both directly (e.g. methanol has two heavy elements, whereas OH and water have only one), and indirectly through changes in the ultraviolet flux, which influences the dissociation rate.  Changes in the relative abundance of different molecular species for similar exciting sources arising in environments with different metallicity would be expected to produce differences in the relative intensities of the associated maser species.  Indeed, it has been speculated that the low numbers of class~II methanol masers in the outer Galaxy, and the relatively lower number of 12.2-GHz methanol masers compared to 6.7-GHz may be due to the lower metallicity \citep{Breen+11}.  This is consistent with the under abundance of methanol masers compared to OH and water masers in low-metallicity local group galaxies, such as the LMC and SMC compared to the Milky Way \citep{Green+08,Ellingsen+10}.  Therefore a maser-based evolutionary scheme is likely to be of limited utility for sources in the outer Galaxy, or extragalactic sources with metallicities significantly different from the Milky Way.  However, the bulk of high-mass star formation within the Milky Way occurs in the mid-plane within the solar circle, where differences in metallicity are relatively small and are hence unlikely to be the limiting factor in the development of a maser-based evolutionary scheme.

The 6.7-GHz transition always shows the largest integrated intensity and velocity range of all the class~II methanol transitions \citep{Breen+11} which demonstrates that it is the most easily inverted.  This interpretation is also consistent with theoretical models \citep{Cragg+05}, and suggests that in general there is a large volume of gas where the 6.7-GHz transition is inverted.  The stochastic nature of the maser process means that of all the class~II methanol maser transitions observed in a region, the parameters of the 6.7-GHz emission will generally be the best guide to the properties of the associated star formation region.  Observations to date of class~II methanol masers are consistent with the basic assumptions that underly the maser-based evolutionary timeline.  For example \citet{Ellingsen07} showed that while the velocity of the peak emission in 6.7-GHz methanol masers often changes on time scales of a decade or longer, the magnitude of the peak flux density typically does not.  Similarly, the correlation between 6.7- and 12.2-GHz methanol maser peak flux densities and the general trend for both the intensity and velocity range of the maser emission to increase with the age of the source observed by \citet{Breen+10a} suggests that the volume of gas contributing to both transitions increases as the source evolves.  The current observations provide additional evidence as well, since given that only the most luminous 12.2-GHz methanol masers have an associated 37.7-GHz methanol maser we might expect there to be a closer relationship between the flux densities of these two transitions than between the 37.7- and 6.7-GHz masers.  However, in direct contrast to this we  found that while there is a correlation between the 6.7- and 37.7-GHz peak flux density ratios with the 6.7- and 12.2-GHz ratio, there is no correlation with the 12.2- and 37.7-GHz flux density ratio (see Section~\ref{sec:complete}).  This suggests that while the properties of the 12.2-GHz methanol masers are a better predictor for the presence and likely peak velocity of 37.7-GHz methanol masers, the intensity of the 37.7-GHz methanol masers is better predicted by the 6.7-GHz properties.

Here, we have used the luminosity of the 6.7- and 12.2-GHz methanol masers to target sources which are likely to have an associated 37.7-GHz methanol maser.  The absence of 37.7-GHz methanol masers in some of these sources may be due to insufficient sensitivity or source variability (e.g. G\,$35.201-1.736$), however, it is also consistent with the 6.7- and 12.2-GHz luminosities being insufficient to identify the specific evolutionary phase associated with this transition.  It is not surprising that the evolutionary phase for a high-mass star formation region cannot be pinpointed through observations of two related maser transitions alone.  The environment where high-mass stars form is complex and their are a large number of potentially important parameters (e.g. stellar mass), and interactions which are unknown, or poorly constrained for many class~II methanol maser sites.  Including observations of other maser species may remove some of the degeneracy present in using only methanol maser observations, for example in section~\ref{sec:OH} we show that the relative intensity of ground-state OH maser emission compared to the 6.7-GHz methanol masers may be a useful to further refine searches for 37.7-GHz methanol masers.   Work is also underway to investigate properties of class~II methanol maser regions through observations of thermal molecular lines and spectral-energy distribution (SED) modelling, as well as through further maser studies.  These observations will provide critical data for testing the assumptions of, and refining the maser-based evolutionary timeline.  

\section{Conclusions}

Our results, in combination with previous searches for 37.7-GHz methanol masers demonstrate that only class~II methanol maser sites with high luminosity in both the 6.7- and 12.2-GHz transitions have an associated 37.7-GHz methanol maser.  The detection rate for 37.7-GHz methanol masers towards sources with a 6.7-GHz methanol maser with an isotropic peak luminosity greater than 800~Jy\,kpc$^2$, and a 12.2-GHz peak luminosity greater than 250~Jy\,kpc$^2$ is at least 30 percent.  The 37.7-GHz methanol masers occur during an evolutionary phase around the time that the associated young high-mass star develops an ultra-compact \ionhy region visible at centimetre wavelengths, and ground-state OH and water masers are present in the majority of these regions.

The 37.7-GHz methanol maser transition shows low-levels of variability on timescales of a few years.  The peak velocity of 37.7-GHz methanol masers aligns with that of the associated 12.2-GHz methanol maser in more than 50 percent of sources and future high spatial and spectral resolution observations of the 37.7-GHz transition offer the prospect of improving existing tests for variations in the proton-to-electron mass ratio at the current epoch.

\section*{Acknowledgements}

We thank an anonymous referee for feedback which has improved this paper.  The Mopra telescope is part of the Australia Telescope which is funded by the Commonwealth of Australia for operation as a National Facility managed by CSIRO.  This research has made use of NASA's Astrophysics Data System Abstract Service.

\bibliography{m405}

\appendix

\section{Additional Evidence for Marginal Detections}
\label{sec:additional}
\begin{figure*}
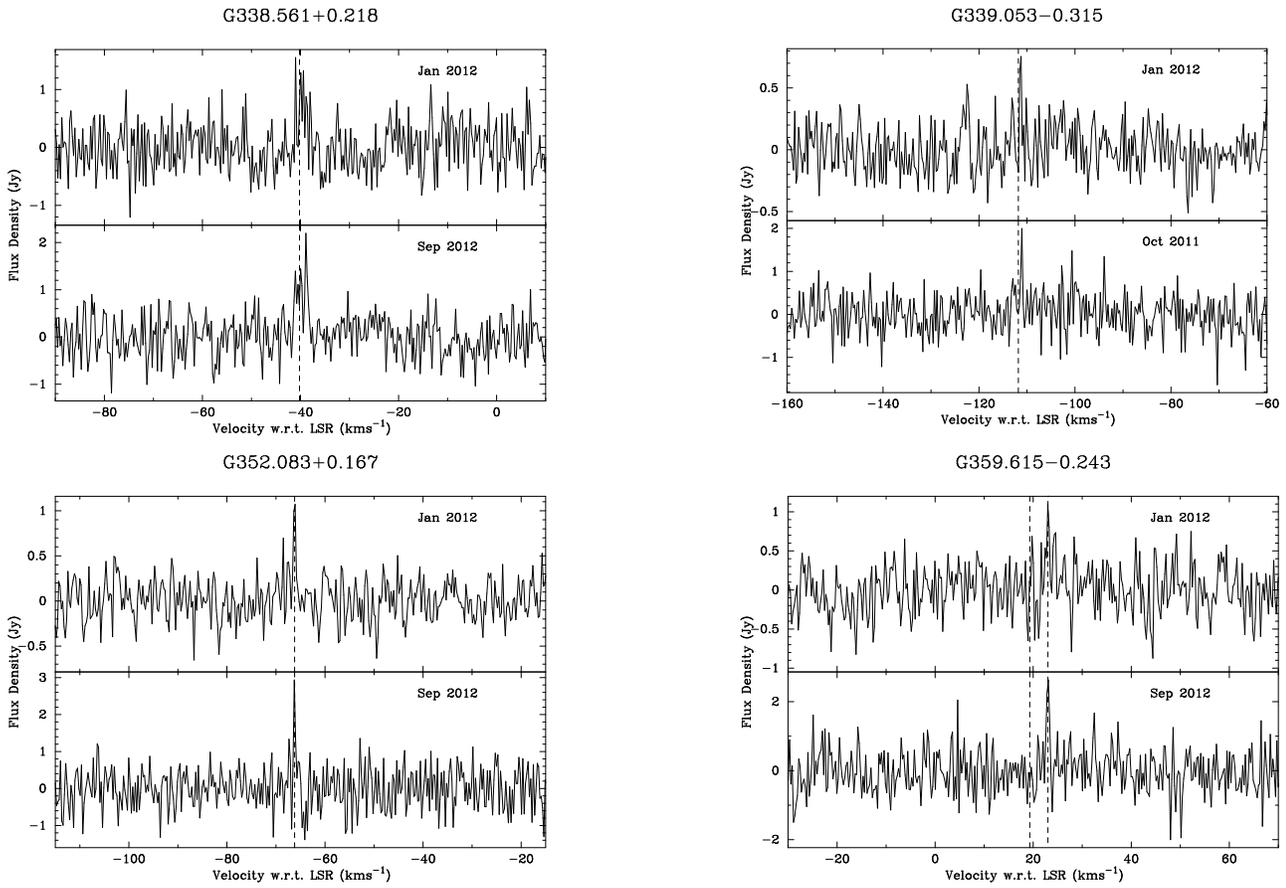

  \specdfig{338.561+0.218_ext}{339.053-0.315_ext}
  \specdfig{352.083+0.167_ext}{359.615-0.243_ext}
   \caption{Spectra of marginal 37.7-GHz methanol masers.  The top spectrum for each source shows the data collected in 2012 January.  The bottom spectrum shows data collected either in 2012 October (for which the flux density calibration is uncertain), or 2011 September.  The dashed vertical line shows the velocity of the 12.2-GHz methanol maser peak observed by \citet{Breen+12a}.  For G\,$359.615-0.243$ the left-hand vertical line is the velocity of the 12.2-GHz peak and the right-hand vertical line is the velocity of the 37.7-GHz peak, which corresponds to a secondary peak in the 12.2-GHz spectrum.}
  \label{fig:evidence}
\end{figure*}

The spectra displayed in Figure~\ref{fig:newdet} for some sources do not meet the typical criterion for being considered a reliable detection (the spectral peak exceeding the mean by more than 5 times the RMS noise).  The spectra exhibit white (Gaussian) noise, so the probability that emission in a single spectral channel will exceed 5-$\sigma$ due to chance is $\sim 2.9 \times 10^{-7}$  (or less than 1 in 3.4 million).  For the data displayed in Figure~\ref{fig:newdet} the peak emission for the sources G\,$338.561-0.218$, G\,$339.053-0.315$, G\,$352.083+0.167$ and G\,$359.615-0.243$ are 4.0, 4.4, 5.1 and 3.5 times the RMS noise respectively.  For all of these sources we have additional data collected with the Mopra telescope either in 2011 October, or 2012 September, which was not included in the spectra shown in Figure~\ref{fig:newdet}.  The setup and strategy for these additional observations was identical to that used in the 2012 January observations, as described in Section~\ref{sec:observations}.  For the observations made in 2012 October (G\,$338.561-0.218$, G\,$352.083+0.167$ and G\,$359.615-0.243$) the flux density scale was incorrect due to problems with the noise diode for the Mopra 7mm receiver at that time.  The data from this session shown in Figure~\ref{fig:evidence} have had the amplitude scale reduced by a factor of three from the nominal calibration, but the flux density scale for these data are likely still not accurate to better than 50 percent.

Figure~\ref{fig:evidence} shows stacked spectra of each of the four marginal 37.7-GHz maser detections.  The top spectrum in each case is the same data presented in Figure~\ref{fig:newdet}, but shown with a larger velocity range, which demonstrates more clearly that the observed peak is higher than the surrounding noise.  The bottom spectrum for each source shows the additional observation, and in each case it can be see that emission with similar signal to noise is observed at the same velocity.  Although the flux density calibration is uncertain for the 2012 October observations, this does not effect the signal to noise ratio (peak flux density divided by the spectrum RMS), which for the additional observations is measured to be 5.5, 4.5, 5.3 and 4.4 for G\,$338.561-0.218$, G\,$339.053-0.315$, G\,$352.083+0.167$ and G\,$359.615-0.243$, respectively.  

If we consider the two sets of observations as independent events, then we have a statistically significant ($>$ 5-$\sigma$) detection for each of the four sources.  For G\,$338.561-0.218$ and G\,$352.083+0.167$, we have a statistically significant detection in one or both of the individual spectra shown in Figure~\ref{fig:evidence}.  The spectra for G\,$339.053-0.315$ show 4.4-$\sigma$ and 4.5-$\sigma$ peaks at -111~\kms\/, in combination this is equivalent to a 6.6-$\sigma$ detection.  The spectra for G\,$359.615-0.243$ show 3.5-$\sigma$ and 4.4-$\sigma$ peaks at 23~\kms\/, in combination this is equivalent to a 5.9-$\sigma$ detection.  The observed 37.7-GHz peak aligns with the peak of the 12.2-GHz emission (dashed vertical line in Fig.~\ref{fig:evidence}) in the same source to within one spectral channel for three of the four cases.  The only exception is G\,$359.615-0.243$, for which the 37.7-GHz peak aligns with a secondary 12.2-GHz spectral feature.  The close alignment of the weak 37.7-GHz methanol masers with the emission observed in other stronger class~II methanol transitions provides additional evidence that these are bona fide detections

\end{document}